\documentclass[11 pt, letterpaper]{article} 
\usepackage[top=1in,bottom=1in,left=1in,right=1in]{geometry} 
\usepackage{amssymb}
\usepackage{times}
\usepackage[belowskip=-15pt,aboveskip=0pt]{caption}

\usepackage{amsmath,amsfonts,bm}

\def\eqref#1{equation~\ref{#1}}

\def\1{\bm{1}}

\DeclareMathAlphabet{\mathsfit}{\encodingdefault}{\sfdefault}{m}{sl}
\SetMathAlphabet{\mathsfit}{bold}{\encodingdefault}{\sfdefault}{bx}{n}

\DeclareMathOperator*{\argmin}{arg\,min}

\usepackage{hyperref}
\usepackage{url}

\usepackage{mathrsfs,dsfont}
\usepackage{amssymb}
\usepackage{xcolor}
\usepackage[english]{babel}
\usepackage[latin1]{inputenc}
\usepackage{color}
\usepackage{bbm}
\usepackage{amsthm}
\usepackage{enumerate}
\RequirePackage{xspace}
\usepackage[ruled, section]{algorithm}
\usepackage{algpseudocode}
\usepackage {graphicx}
\usepackage {pgf}
\usepackage{yhmath}
\usepackage[normalem]{ulem}
\usepackage{multido}
\usepackage{pstricks,pst-plot,pstricks-add,pst-math}
\usepackage{mathtools}
\usepackage{stmaryrd}
\usepackage{caption}
\usepackage{subcaption}

\numberwithin{equation}{section}
\numberwithin{figure}{section}
\numberwithin{table}{section}
\theoremstyle{plain}

\theoremstyle{definition}

\theoremstyle{remark}

\renewcommand{\ge}{\geqslant}

\def\beq{ \begin{equation} }
\def\eeq{ \end{equation} }
\def\beqx{ \begin{equation*} }
\def\eeqx{ \end{equation*} }
\def\beqa{\begin{eqnarray}}
\def\eeqa{\end{eqnarray}}
\def\beqax{\begin{eqnarray*}}
	\def\eeqax{\end{eqnarray*}}

\usepackage[T3,T1]{fontenc}
\DeclareSymbolFont{tipa}{T3}{cmr}{m}{n}
\DeclareMathAccent{\inv}{\mathalpha}{tipa}{16}
\allowdisplaybreaks

\usepackage[round]{natbib}
\renewcommand{\bibname}{References}
\renewcommand{\bibsection}{\subsubsection*{\bibname}}

\title{Estimating the treatment effect of the juvenile stay-at-home order on SARS-CoV-2 infection spread in Saline County, Arkansas}

\author{Neil Hwang, Shirshendu Chatterjee \\
Department of Mathematics, City University of New York\\
Yanming Di \& Sharmodeep Bhattacharyya \thanks{Corresponding Email:  bhattash@science.oregonstate.edu.}\\
Department of Statistics, Oregon State University \\
}

\begin{document}

\maketitle

\begin{abstract}
  We investigate the treatment effect of the juvenile stay-at-home order (JSAHO) adopted in Saline County, Arkansas, from April 6 to May 7, in mitigating the growth of SARS-CoV-2 infection rates. To estimate the counterfactual control outcome for Saline County, we apply Difference-in-Differences and Synthetic Control design methodologies. Both approaches show that stay-at-home order (SAHO) significantly reduced the growth rate of the infections in Saline County during the period the policy was in effect, contrary to some of the findings in the literature that cast doubt on general causal impact of SAHO with narrower scopes. 
\end{abstract}

\section{Introduction}

In response to rising numbers of Covid-19 cases, state governments have implemented a wide-ranging array of policies in the form of non-pharmaceutical interventions (NPIs) aimed at slowing the rate of growth of the SARS-CoV-2 infections. At times, local governments have stepped up and issued orders when the policies they deemed necessary were not implemented at the state level. Much of this policy response has focused on enforcing social-distancing through measures ranging from temporary closures of public-facing businesses and shelter-in-place orders (SIPO) to mandatory mask-wearing orders [\citet{abouk2020immediate, friedson2020did, courtemanche2020strong, chernozhukov2020causal, dave2020shelter, hsiang2020effect}]. The prominence of state and local governments in implementing policy responses v\'{i}s-a-v\'{i}s the federal government is explained by the fact that the jurisdictional authority to do so rests with the former [\citet{dave2020shelter}]. 
  
Of the various policies that have been implemented, SIPO has been found to be among the most effective in slowing the growth of infections in the U.S. There have been a plethora of studies on association between SIPO and SAHO and SARS-CoV-2 infection spread [\citet{gao2020association, le2020impact, lurie2020covid}].  \citet{abouk2020immediate} used the difference-in-differences methodology to conclude that statewide stay-at-home orders (SAHO) showed the strongest causal effect on reducing social interactions at the state-level [\citet{abouk2020immediate}]. \citet{courtemanche2020strong} noted significant causal impact of interventions, including SIPO, in reducing case growth rates at the county-level [\citet{courtemanche2020strong}]. \citet{chen2020causal} also investigates the causal impact of SAHO on mobility and SARS-CoV-2 infection spread. Regarding SIPO, \citet{dave2020shelter} noted that while its effectiveness is most notable when adopted in dense areas early in the pandemic, its impact declines significantly when implemented later [\citet{dave2020shelter}]. \citet{friedson2020did} used synthetic control research design to find that SIPO in California reduced both cases and deaths. It was also noted that SIPO is the most restrictive form of social-distancing measure with its compliance assurance coming from law enforcement and punitive fines [\citet{dave2020shelter, friedson2020did, caswell,napoleon_2020}], as well as social pressures [\citet{dave2020shelter,ronayne}]. 

A similar but more lenient "advisory" SIPO or stay-at-home order (SAHO) discourages members of the public from leaving their homes other than for medical emergencies, commuting to work, or shopping for necessities. SAHO with narrower scopes have also been issued, pertaining to only a certain segment of the resident population. Juvenile stay-at-home orders (JSAHO), such as that issued in Saline County of Arkansas [\citet{salinecountyorder}], comprise one such example and allow juveniles to leave their homes if accompanied by an adult. However, strict adherence to such measures is often not enforced; instead, its effectiveness to a large extent relies on the public's willingness to modify their behaviors to comply in light of the pandemic and SAHO [\citet{salinecountyorder}]. \citet{abouk2020immediate} found that SAHOs with narrower scopes did not show significant causal effect in reducing infection rates at the state level [\citet{abouk2020immediate}]. Then, the natural question arises as to how effective such measures are that rely, at least in part, on the public to voluntarily comply.

We answer this question by examining JSAHO that Saline County in Arkansas adopted as a policy response to the pandemic and the nature of its impact, if any, while the policy was in effect. As one of five states that did not impose either SIPO or SAHO as of September 5, 2020, Arkansas is unique in that one of its counties, Saline County, nonetheless issued its own SAHO with the scope restricted to those under the age of 18. This provides an invaluable setting for natural experiments to assess causal effects of policy actions adopted at the county level, such as JSAHO, while controlling the effects of other policies that have been concurrently implemented at the state-level. In particular, examining counties in one state allows us to control for the effect of the virus testing on cases since the availability of tests is constant at the state level.

A number of methodologies have been used in the Covid-19 policy evaluation literature to estimate average treatment effects using panel data. Among the more popular methodologies include Difference-in-Differences (DID) [\citet{abouk2020immediate}], propensity scores matching (PSM) [\citet{courtemanche2020strong}], event-studies [\citet{abouk2020immediate,courtemanche2020strong, dave2020shelter}], and synthetic control research designs (SC) [\citet{friedson2020did,abadie2003economic,abadie2005semiparametric,abadie2010synthetic,abadie2015comparative}]. Of these methods, SC is a relatively novel method in the broader econometrics literature and has gained an increasingly larger following in recent years, spawning several varieties and enhancements in the process. 

The rest of the paper is organized as follows. In Section \ref{DIDnotation}, we state the terms and notations used throughout this paper. In Section \ref{DID}, we provide an overview of the standard DID methodology, apply it empirically to the Covid-19 data in Arkansas, discuss the results, and note key shortcomings of the approach. Similarly, in Section \ref{SC}, we introduce the SC methodology in its standard form and discuss empirical findings resulting from its application to the data. In Section \ref{data_sources}, we reference the sources of the data used for empirical analyses in this paper. We conclude with some suggestions for further study on both the empirical and methodological fronts. Lastly, various visualizations and regression output are shown in the Appendix.

\section{Notation}
\label{DIDnotation}
As is standard in the literature on policy evaluation, we use the term "treatment" to specifically refer to JSAHO and "treatment group" to refer to Saline County, AK, since it was the only county in which residents received the treatment. Similarly, "control group" refers to some set of counties in Arkansas other than Saline County, where individual counties of the control group are referred to as "control units." 

We denote infection rates in county $i$ belonging to group g $\in \{1($treatment$),0($control$)\}$ in Arkansas at time $t$ by $Y^g_{it}$. As noted earlier, the only county in the treatment group in our study is Saline County, and the membership composition of the control group would vary depending on the methodology under study, as discussed below. Without loss of generality, we assign Saline County the index $i=1$, and denote its infection rate by $Y^1_{1t}$, and the counties in the control group $Y^0_{jt}$, with $j\in [2,75]$. The time variable is binary, $0$ for pre-treatment period, i.e., before \textbf{JSAHO was issued on April 6}, and $1$ for treatment period, i.e., \textbf{April 6 through May 7} when the order was lifted in Saline County. The days after May 7 are referred to as the post-treatment period. 

$K$ denotes the number of pertinent covariates included in a given model to explain the variation in pre-treatment infection rates of Saline County and $N$ refers to the number of control units. Then, $\textbf{X}_0$ is a $K\times N$ matrix containing the values of the covariates for $N$ control units. For the treatment unit (Saline County), we denote by $\textbf{X}_1$ a $K\times 1$ vector of pre-treatment values of $K$ covariates. 

\section{Difference-in-Differences (DID)}\label{DID}
Also known as a methodology for "natural" or "quasi-experiments," DID has been a popular methodology of choice among applied researchers in policy analysis for its simplicity and intuitive appeal. In particular, the JSAHO setting in Arkansas lends itself well for DID analysis. We first discuss the DID methodology in light of the pandemic setting, followed by our empirical study design and findings.

\subsection{Methodology}
The estimate for the treatment effect is simply the difference between mean reduction in infection rate in the treatment county and that in a control county, where this double differencing is meant to remove biases due to county fixed effects and time effects. 

In its canonical form, the estimand for the treatment effect of the policy, denoted $\tau$, is expressed in turn as follows:
\begin{align*}
	\tau &= \mathbb{E}[Y^1_{11} - Y^1_{10}] - \mathbb{E}[Y^0_{\cdot1}-Y^0_{\cdot0}]
\end{align*}
The foremost assumption in DID (known as the parallel trends assumption) is that the time trends for both groups are identical, and hence the subtraction of the two expectations is designed to eliminate the common time effect. With this assumption, the DID estimate of the treatment effect is unbiased. 

$\tau$ is typically estimated fitting a linear regression line of the following form [\cite{ashenfelter1984using}]:
\begin{equation}\label{DIDregression}
	Y^g_{it}= \alpha + \beta'\cdot t_i + \gamma_i\cdot g_i + \tau \cdot g_i t_i +\varepsilon_i
\end{equation}
where $\alpha, \beta,$ and $\gamma$ are model parameters for the intercept, the time coefficient applicable to both groups, and the group-specific coefficient, respectively. Finally, $\varepsilon_i$ represents unobservable traits of county $i$, and are assumed to be 
independent of $g_i$ and $t_i$, and that $\mathbb{E}[\varepsilon_i]=0$.

The estimate for $\tau$ is thus
\begin{align*}
\hat{\tau} &= (\bar{Y}^1_{11}-\bar{Y}^1_{10})-(\bar{Y}^0_{\cdot 1}-\bar{Y}^0_{\cdot 0})\\
&= \big(Y_{11} - 
Y_{10} \big) 
- \bigg(\sum_{i|g_i = 0, t_i = 1} \frac{Y_{i}}{|i|g_i = 0, t_i = 1|} - 
\sum_{i|g_i = 0, t_i = 0} \frac{Y_{i}}{|i|g_i = 0, t_i = 0|} \bigg)
\end{align*}

\subsection{Discussion}
As stated earlier, the most crucial assumption in the DID design is that the groups exhibit parallel trends over time so as to allow for the cancellation of the time effects in differencing the differences. This assumption is crucial since one does not observe the counterfactual infection rate of the treatment unit in absence of the treatment $Y^0_{11}$, i.e., the infection rate that Saline County would have had it not implemented JHAHO. The parallel trends assumption allows us to infer the counterfactual outcome based on $Y^1_{10}$, $Y^0_{i0}$, and $Y^0_{i1}$ with $g_i=0$ and $i\in [2,75]$.

For this reason, a standard approach is to choose as a control a county that has the pre-treatment period time trend similar to that of Saline County, where the pre-treatment period starts from the date the first case was reported until the day before the start of the JSAHO effective dates. More concretely, one would choose a county that exhibits the slope coefficient that is close to that of Saline. However, one cautionary note regarding time trends in the pre-treatment period is that control units typically exhibit multiple time trends in the pre-treatment period. 

\begin{figure}[H]
	\centering
	\begin{subfigure}[b]{0.45\textwidth}
		\centering
    	\includegraphics[width = 0.7\textwidth]{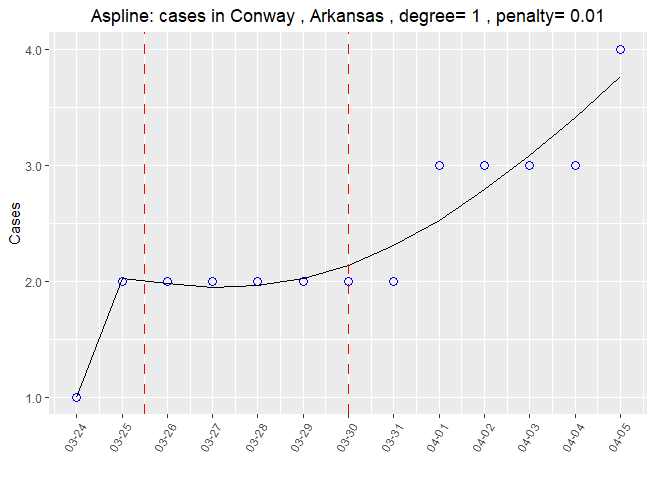}
    	\caption{Cumulative Cases in Conway County}
    	\label{conway_knots}
	\end{subfigure}
\begin{subfigure}[b]{0.45\textwidth}
	\centering
	\includegraphics[width = 0.7\textwidth]{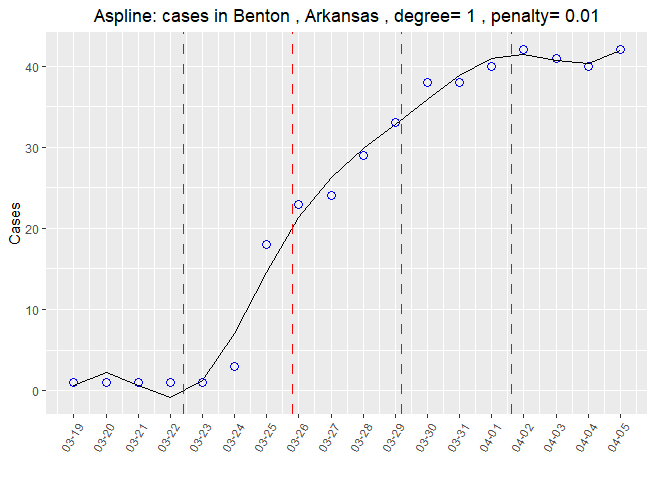}
	\caption{Cumulative Cases in Benton County}
	\label{benton_knots}
\end{subfigure}
\end{figure}

For example, Figures \ref{conway_knots} and \ref{benton_knots} show the time series plots of cumulative case counts in Conway and Benton Counties. The dotted red lines represent the dates when the slopes of the fitted curves change, also known as "change-points" or "knots." 

It is obvious from the figures above that the time trends for both counties change depending on the specific date ranges one considers. 

For this reason, when selecting candidate counties for control units, we first identified for each county change-points where the changes in the slopes of fitted lines were significant. We used the knot-selection algorithm in the adaptive splines method discussed in \cite{goepp2018spline} and implemented in the \textit{A-Splines} R package.

\begin{figure}[H]
	\centering
	\includegraphics[width = 0.8\textwidth]{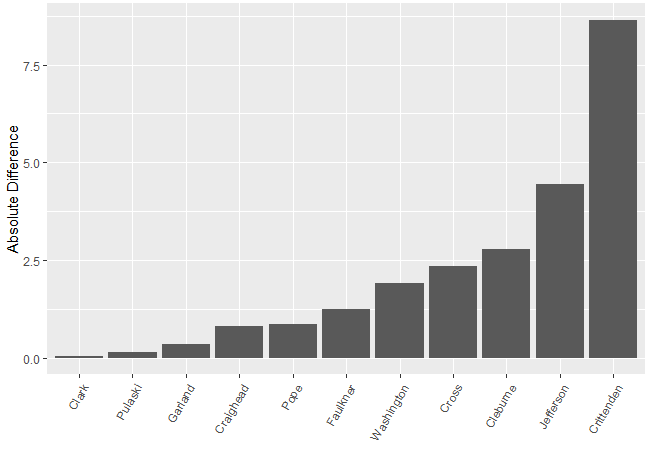}
	\caption{Counties by absolute deviations of slope coefficients vs. Saline County}
	\label{slope_coefficient}
\end{figure}

Then, we considered only the dates after the most recent change-point when fitting a line for each county to estimate the time trend. We selected the counties that had slope coefficients significant at the 5\% level with adjusted $R$-squared of 0.75. Figure \ref{slope_coefficient} shows 11 counties obtained as a result of this procedure and constitute potential control units listed in order of the absolute deviation of the slope coefficient from that of Saline County. 

\begin{figure}[H]
	\centering
	\begin{subfigure}[b]{0.45\textwidth}
		\centering
    	\includegraphics[width = 1.0\textwidth]{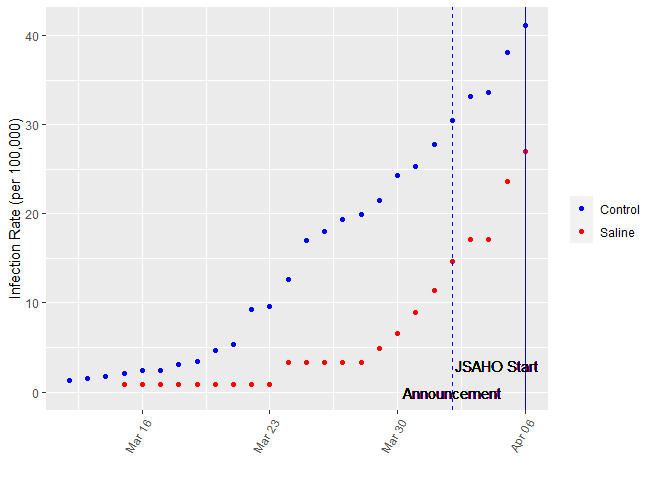}
    	\caption{Counties with < 1.0 Abs. Deviation as Control}
    	\label{saline_top5}
	\end{subfigure}
\begin{subfigure}[b]{0.45\textwidth}
	\centering
	\includegraphics[width = 1.0\textwidth]{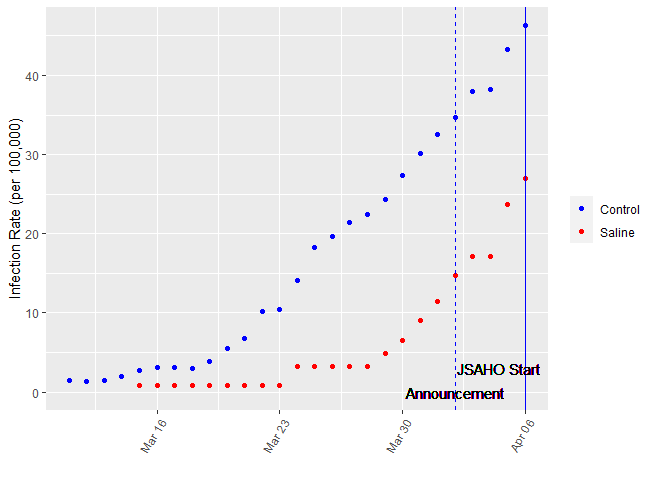}
	\caption{All Potential Counties as Control}
	\label{saline_all}
\end{subfigure}
\end{figure}

Selecting those counties with the absolute difference less than 1.0 to comprise the control group, its mean infection rates versus those of the treatment unit as shown in Figure \ref{saline_top5}. Selecting all the counties in Figure \ref{slope_coefficient} as the control group produces the mean infection rate plot shown in Figure \ref{saline_all}. In both cases, one can visually confirm that the time trends of the infection rates for both the treatment and control groups are very similar heading into the treatment period beginning date of April 6. 

One may wonder how the control units compare to the treatment on pertinent covariates. In the Appendix are figures that show how the treatment and each control unit compare on various social and demographic indicators. For instance, in Figure \ref{covariates_saline_pulaski} the treatment unit is more dense and has lower mortality rates, although whether the latter affects the infection rates is uncertain. The control unit has a lower median age. In Figure \ref{covariates_saline_garland}, the density appears to be about the same for both counties. The treatment unit has lower mortality and poverty rates. 

When considering each potential control unit separate, we found that the treatment effect was significant when using 7 of the 11 control units. To illustrate, we first consider 2 such control units, Pulaski and Garland. Both counties had relatively small absolute deviations in pre-treatment trends and had similar values for several key covariates in comparison to Saline County. Then, we assess results based on the other remaining control units. 

Figures \ref{saline_pulaski} and \ref{saline_garland} below show the infection rate trends of Saline vs. Pulaski and Garland counties juxtaposed in the pre-treatment period. The blue dotted line indicates the date when JSAHO was issued on April 2 and the solid blue line indicates the start of the effective dates of April 6. One can confirm that the time trends starting on March 30 and onwards for the treatment and control units are very similar in both graphs. Indeed, based on the infection rates between the last change-point (March 29 for Saline and Pulaski, and March 30 for Garland) and the start of the treatment period, the slope coefficients are 2.52, 2.38, and 2.17 for Saline, Pulaski, and Garland counties, respectively.

\begin{figure}[H]
	\centering
	\begin{subfigure}[b]{0.45\textwidth}
		\centering
    	\includegraphics[width = 0.9\textwidth]{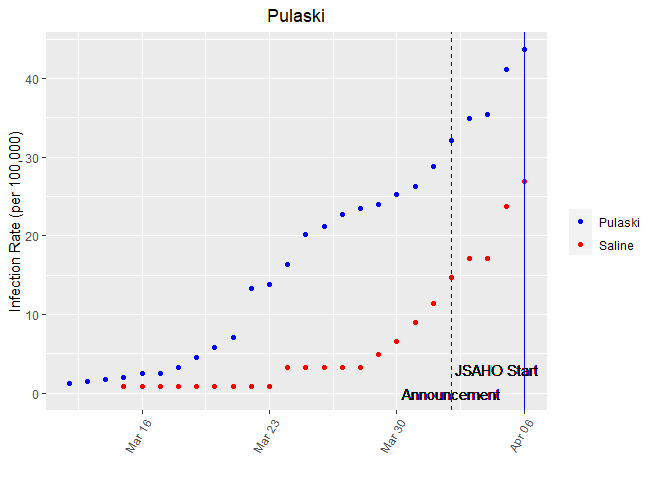}
    	\caption{Infection Rates of Saline and Pulaski}
    	\label{saline_pulaski}
	\end{subfigure}
\begin{subfigure}[b]{0.45\textwidth}
	\centering
	\includegraphics[width = 0.9\textwidth]{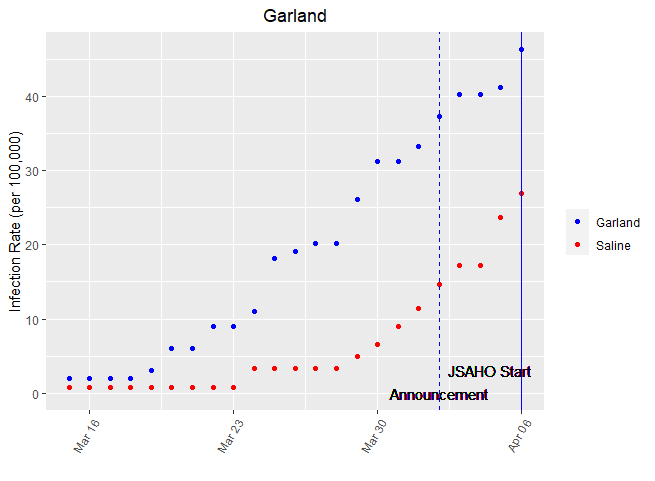}
	\caption{Infection Rates of Saline and Garland}
	\label{saline_garland}
\end{subfigure}
\end{figure}

\vspace{1cm}

Lastly, as noted in \cite{bertrand2004much}, conventional standard error estimates for the treatment effect using the OLS (\ref{DIDregression}) often suffer from downward bias due to serial correlation in infection rates within each county. Prior works in the literature have addressed this issue by clustering standard errors at the geographical level where measurements are taken [\cite{chernozhukov2020causal}, \cite{abouk2020immediate}]. For this reason, we cluster standard errors at the county level. 

\subsection{Empirical Results}

It has been documented that the incubation period of the Coronavirus ranges from 2 to 14 days, with the median of 5 days [\citet{guan2020clinical, lauer2020incubation,dave2020shelter}], while others have noted that the effect of policies is likely to be observed with delay [\citet{abouk2020immediate, dave2020shelter}]. For this reason, to fully assess the causal impact of the policy, we examine infection rates through the end of the policy treatment period plus 7 days. Figures show the progressions of infection rates for the treatment in red and the two control groups units in blue. The black vertical line indicates May 7 when JSAHO was lifted. 

\begin{figure}[H]
	\centering
	\begin{subfigure}[b]{0.45\textwidth}
		\centering
    	\includegraphics[width = 1.0\textwidth]{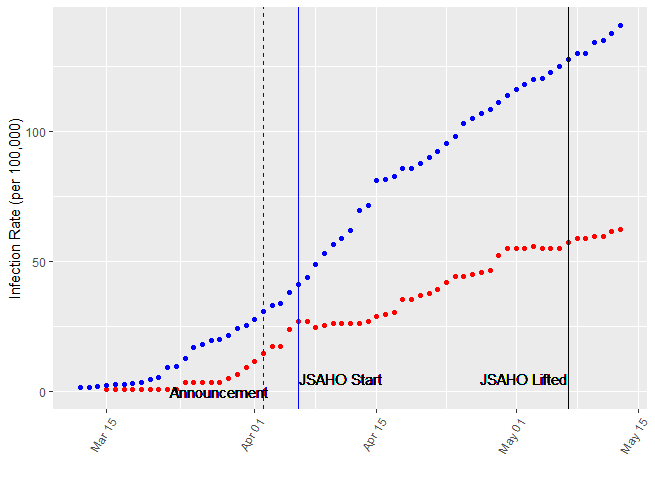}
    	\caption{Counties with < 1.0 Absolute Difference}
    	\label{treatment_saline_top5}
	\end{subfigure}
\begin{subfigure}[b]{0.45\textwidth}
	\centering
	\includegraphics[width = 1.0\textwidth]{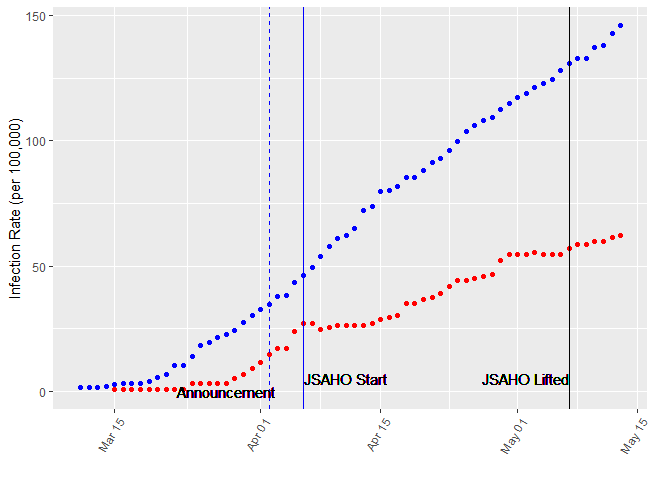}
	\caption{Saline vs. All Candidate Control Counties}
	\label{treatment_saline_allcontrol}
\end{subfigure}
\end{figure}

Figures \ref{treatment_saline_top5} and \ref{treatment_saline_allcontrol} show the the mean infection rates of the 5-county and 11-county control groups compared to that of the treatment unit over the pre-treatment and treatment periods, plus 7 days. 

\vspace{1cm}

\begin{figure}[H]
	\centering
	\includegraphics[width = .7\textwidth]{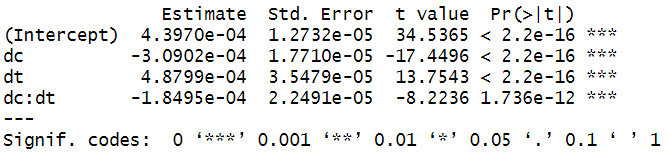}
	\caption{Counties with Absolute Difference < 1.0}
	\label{lm_model_saline_top5}
\end{figure}

\begin{figure}[H]
	\centering
	\includegraphics[width = 0.7\textwidth]{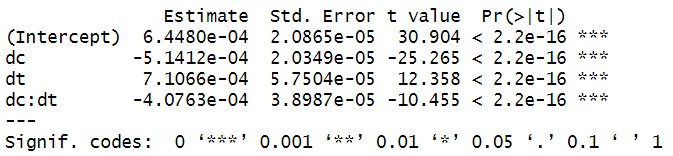}
	\caption{All Potential Control Units}
	\label{lm_model_saline_allcontrol}
\end{figure}

Visually, one can confirm that there is a relative reduction in the growth rate of infections in Saline during the treatment period in both cases. In both cases, \ref{treatment_saline_top5}, the downward effect on the infection rates appears to continue after May 7, the last day of the treatment period, which is consistent with observations made in prior works regarding the lag in policy effectiveness [\cite{abouk2020immediate}]. To estimate the treatment effect parameter, we fit Equation \ref{DIDregression} to assess the value of the coefficient $\tau$. As shown in Figures \ref{lm_model_saline_top5} and \ref{lm_model_saline_allcontrol}, the estimand for $\tau$ is the coeffient for the parameter $dc:dt$, where $dc$ is an indicator for the county fixed effects and $dt$ is an indicator for the time effects. The coefficient for $dc:dt$ is $-1.85\times 10^{-0.4}$ with the 5-county control group and $-4.08\times 10^{-0.4}$ with all 11 counties comprising the control group. Both estimates are shown to be significant based on clustered standard errors at the county level. 
\vspace{0.5cm}

\vspace{1cm}

We also observe similar findings when examining individual counties as the control group. For example, below are the infection rate trends for the treatment versus Pulaski and Garland counties as controls. 

\begin{figure}[H]
	\centering
	\begin{subfigure}[b]{0.45\textwidth}
		\centering
    	\includegraphics[width = 1.0\textwidth]{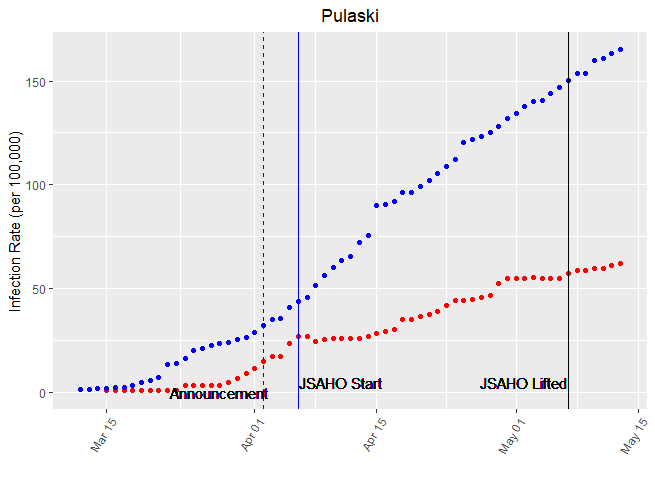}
    	\caption{Saline vs. Pulaski}
    	\label{treatment_saline_pulaski}
	\end{subfigure}
\begin{subfigure}[b]{0.45\textwidth}
	\centering
	\includegraphics[width = 1.0\textwidth]{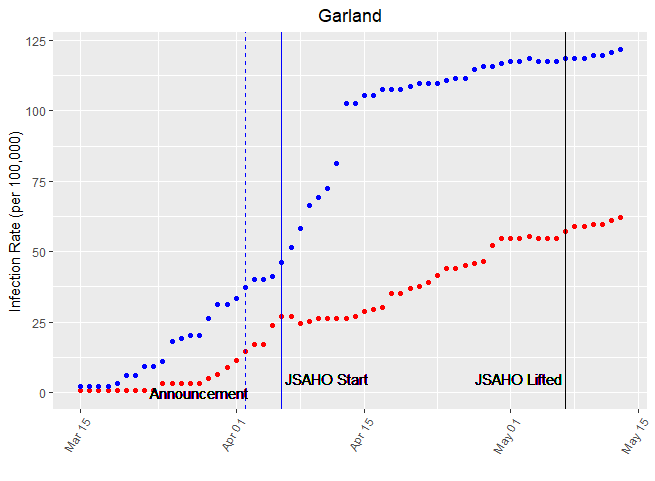}
	\caption{Saline vs. Garland}
	\label{treatment_saline_garland}
\end{subfigure}
\end{figure}

Figures \ref{treatment_saline_pulaski} and \ref{treatment_saline_garland} show the infection rates of Pulaski and Garland counties as the control units compared to that of the treatment unit. 

\begin{figure}[!ht]
	\centering
	\includegraphics[width = .7\textwidth]{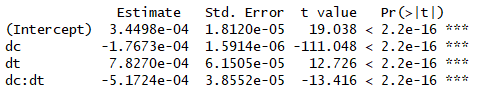}
	\caption{Treatment Effect Estimate with Pulaski as Control}
	\label{lm_model_saline_pulaski}
\end{figure}

\begin{figure}[!ht]
	\centering
	\includegraphics[width = .7\textwidth]{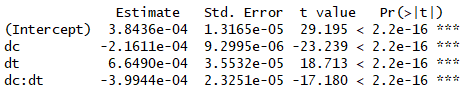}
	\caption{Treatment Effect Estimate with Garland as Control}
	\label{lm_model_saline_garland}
\end{figure}

As before, there is a reduction in the infection rates in the treatment during the treatment period in both cases, with the policy lag exhibiting in the post-treatment period shown in Figure \ref{treatment_saline_pulaski}. Figures \ref{lm_model_saline_pulaski} and \ref{lm_model_saline_garland} show the estimands for $\tau$: $-5.17\times 10^{-0.4}$ with Pulaski as the control and $-3.99\times 10^{-0.4}$ with Garland as the control, both of which are significant based on clustered standard errors at the county level. 

\begin{figure}[H]
	\centering
	\includegraphics[width = .7\textwidth]{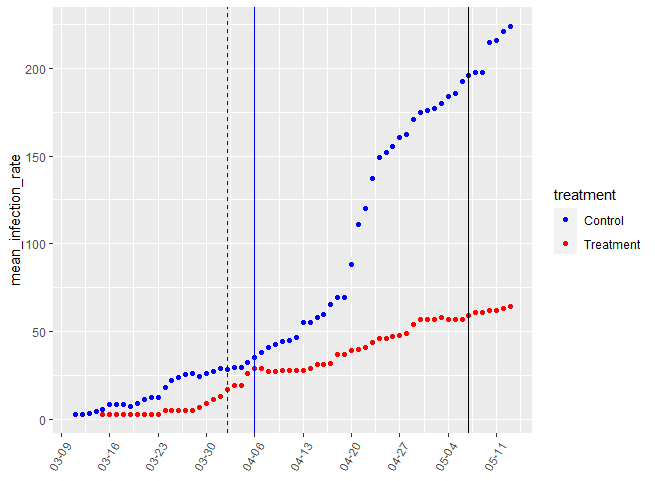}
	\caption{Infection Rates of Saline vs. All the Other Counties}
	\label{treatment_saline_all}
\end{figure}

\vspace{1cm}
As a reference, one can compare the infection rate of Saline County and the mean infection rates of the other 74 counties in Arkansas, as shown in Figure \ref{treatment_saline_all}. 

Interestingly, the time trends in the pre-treatment period are not parallel, with Saline County showing a steeper slope than the mean of the control counties' trends. However, the reduction in infection rate for Saline County v\'{i}s-a-v\'{i}s the control is nonetheless obvious. 

\begin{figure}[H]
	\centering
	\includegraphics[width = 1\textwidth]{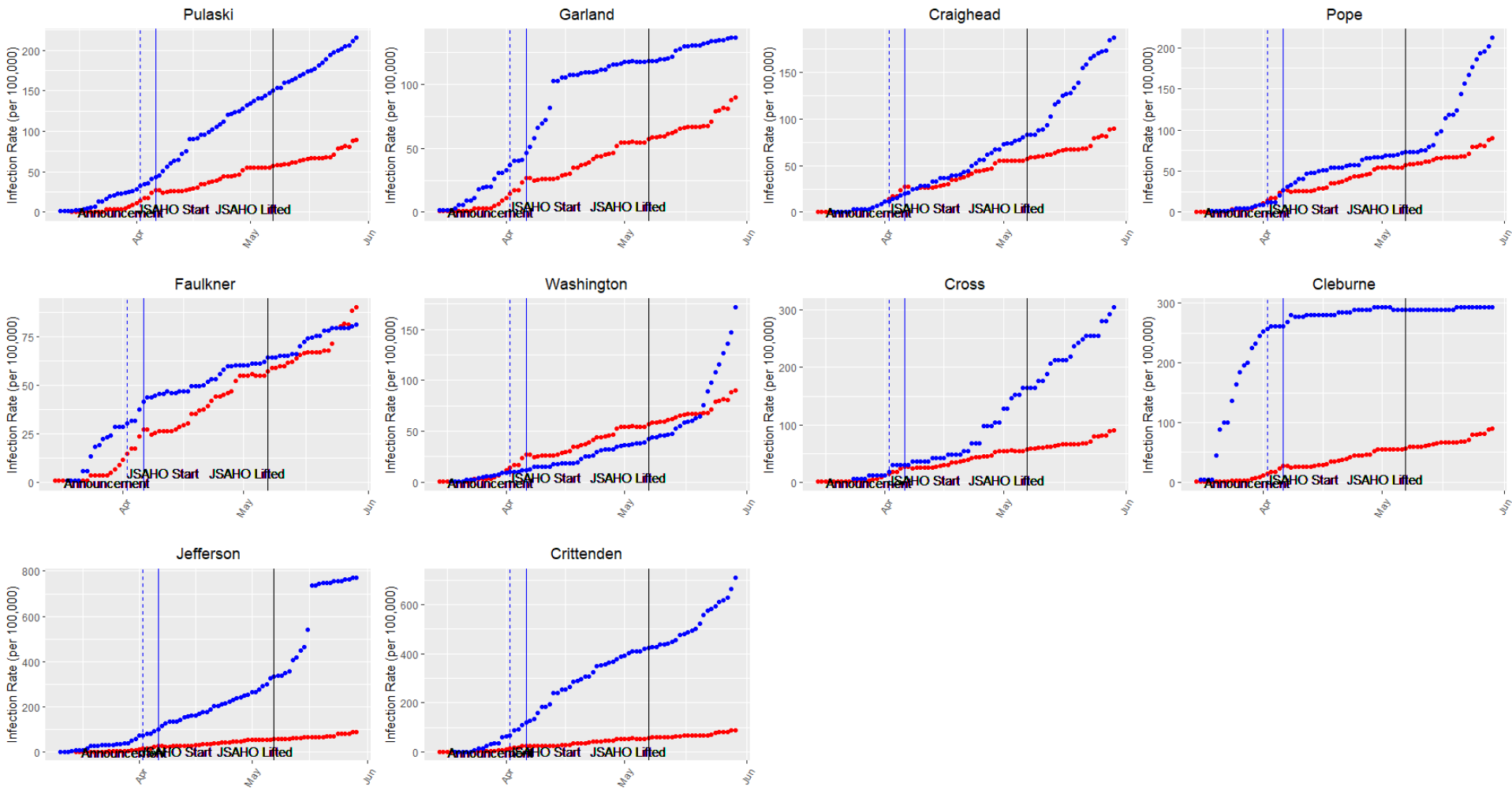}
	\caption{Infection rates for Saline (red) vs. the Remaining Counties as the Control (blue)}
	\label{all_control_plots}
\end{figure}

Figure \ref{all_control_plots} shows the comparative time series of infection rates for all control counties except Clark county. One can visually confirm the presence of the treatment effect in 5 of the 8 remaining counties, namely, Craighead, Pope, Cross, Jefferson, and Crittenden. 

\begin{figure}[H]
	\centering
	\includegraphics[width = 1\textwidth]{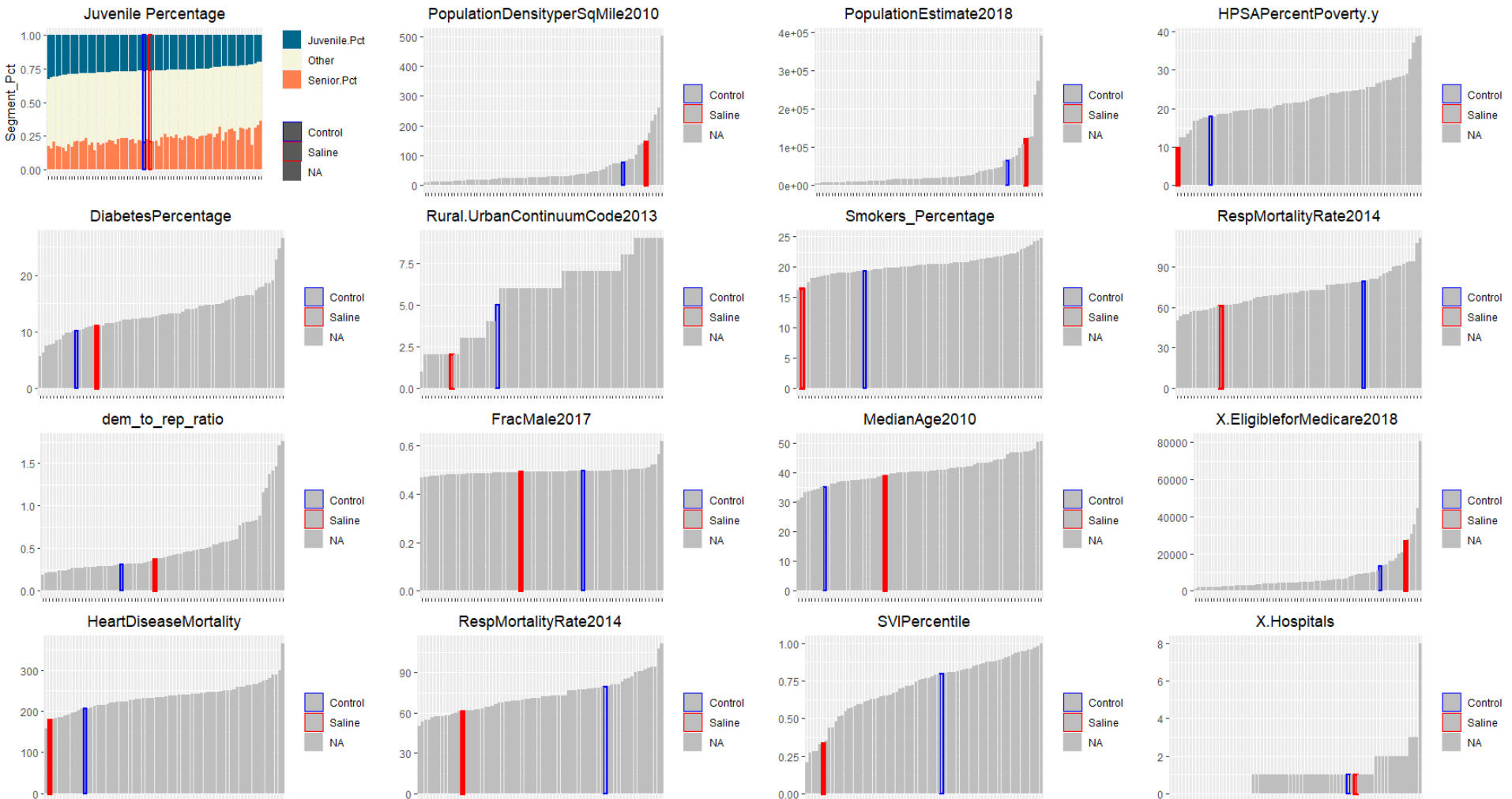}
	\caption{Covariate Values of Saline and Clark Counties}
	\label{covariates_saline_clark}
\end{figure}

\begin{figure}[H]
	\centering
	\includegraphics[width = 1\textwidth]{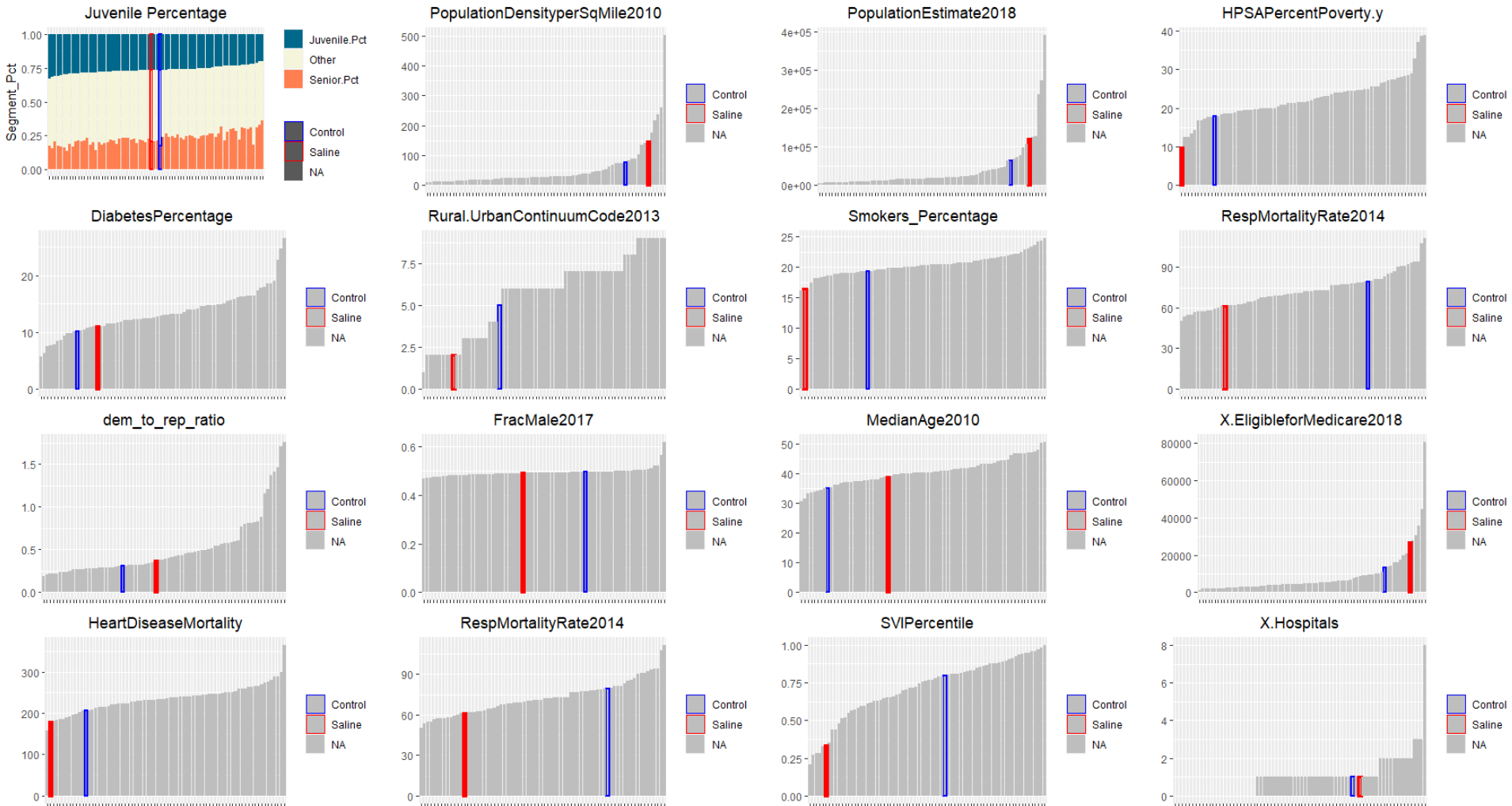}
	\caption{Covariate Values of Saline vs. Pulaski Counties}
	\label{covariates_saline_pulaski}
\end{figure}

\begin{figure}[H]
	\centering
	\includegraphics[width = 1\textwidth]{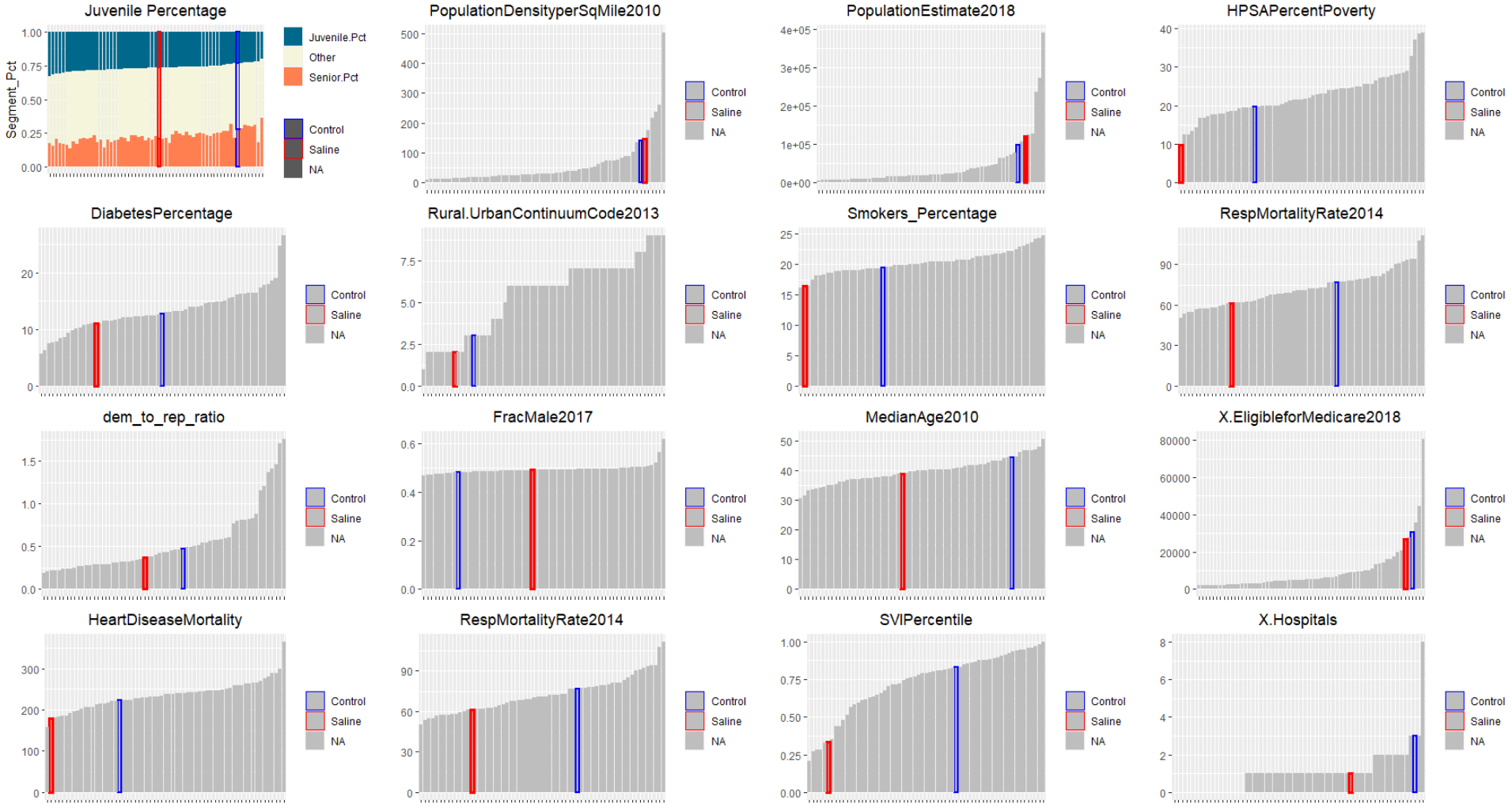}
	\caption{Covariate Values of Saline vs. Garland Counties}
	\label{covariates_saline_garland}
\end{figure}

\begin{figure}[H]
	\centering
	\includegraphics[width = 1\textwidth]{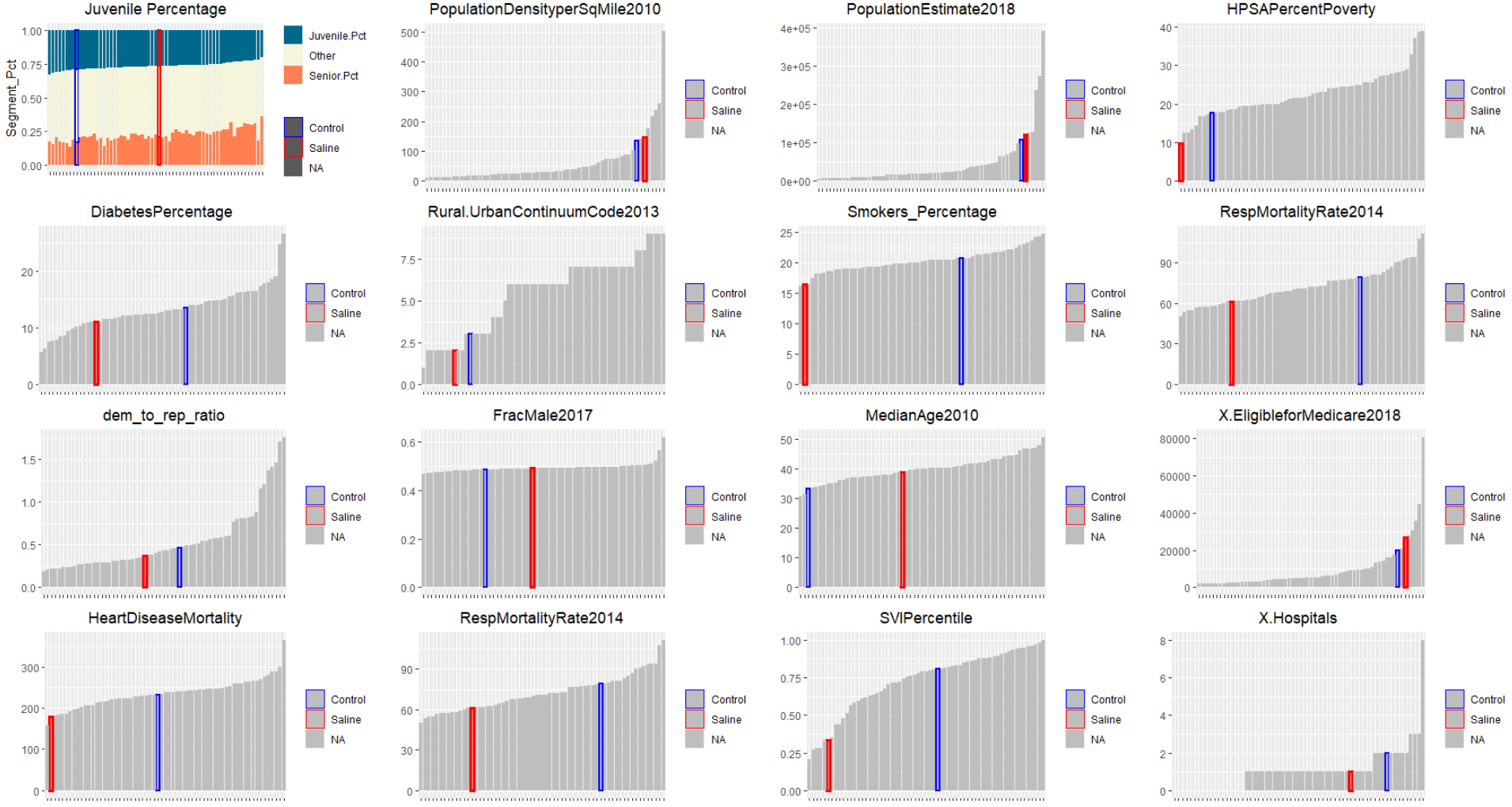}
	\caption{Covariate Values of Saline vs. Craighead Counties}
	\label{covariates_saline_craighead}
\end{figure}

\begin{figure}[H]
	\centering
	\includegraphics[width = 1\textwidth]{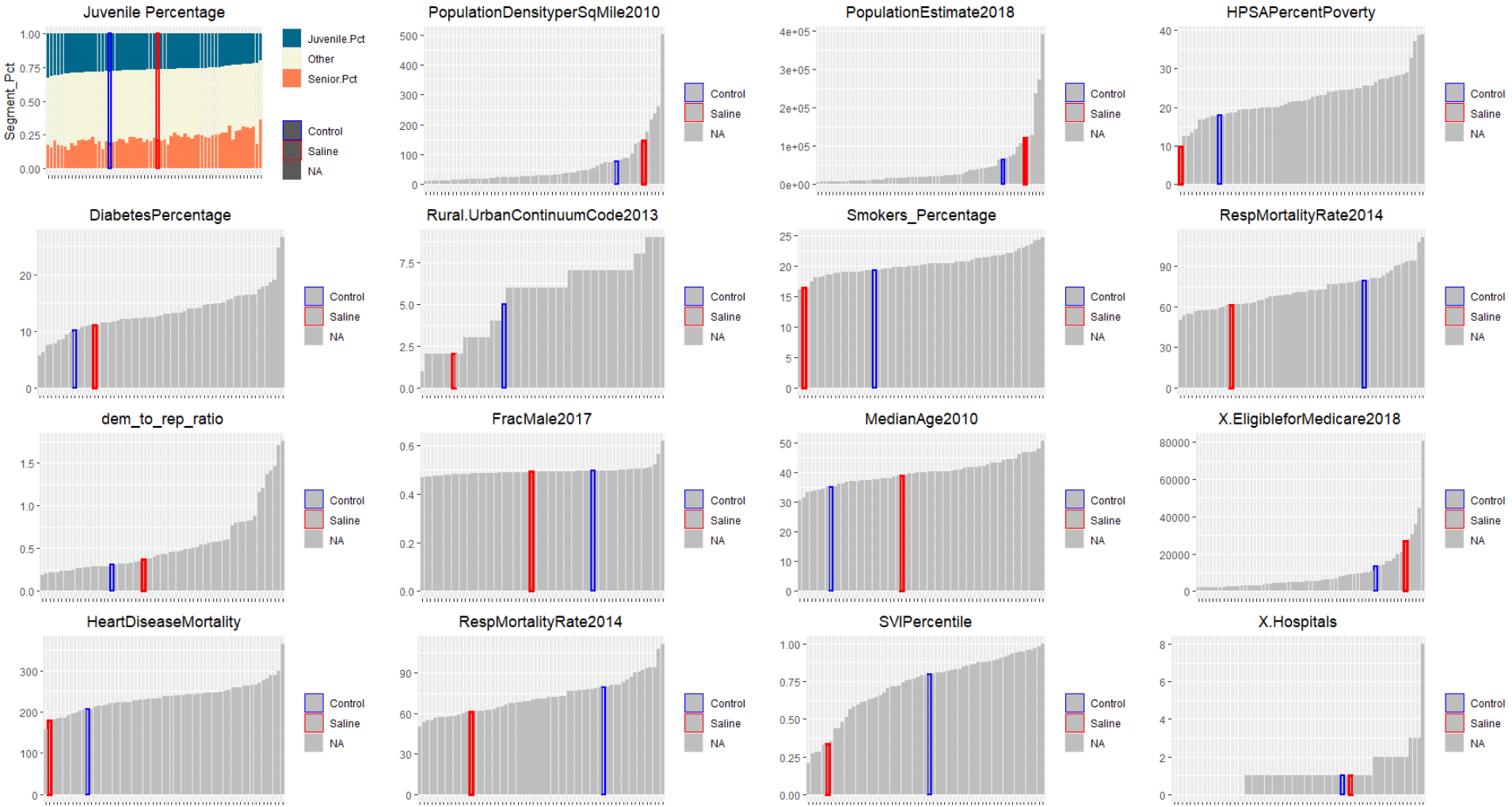}
	\caption{Covariate Values of Saline vs. Pope Counties}
	\label{covariates_saline_pope}
\end{figure}

\begin{figure}[H]
	\centering
	\includegraphics[width = 1\textwidth]{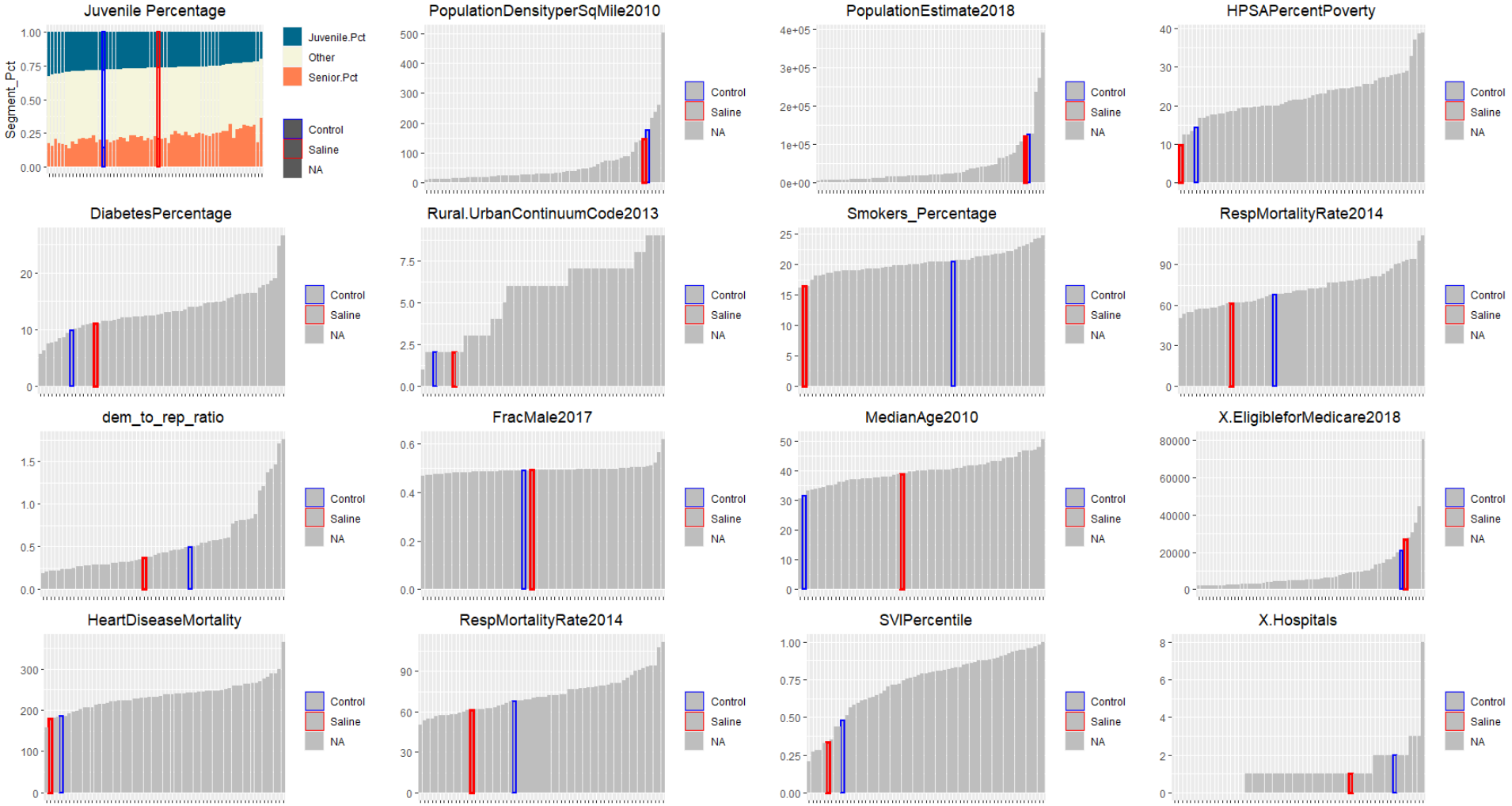}
	\caption{Covariate Values of Saline vs. Faulkner Counties}
	\label{covariates_saline_faulkner}
\end{figure}

\begin{figure}[H]
	\centering
	\includegraphics[width = 1\textwidth]{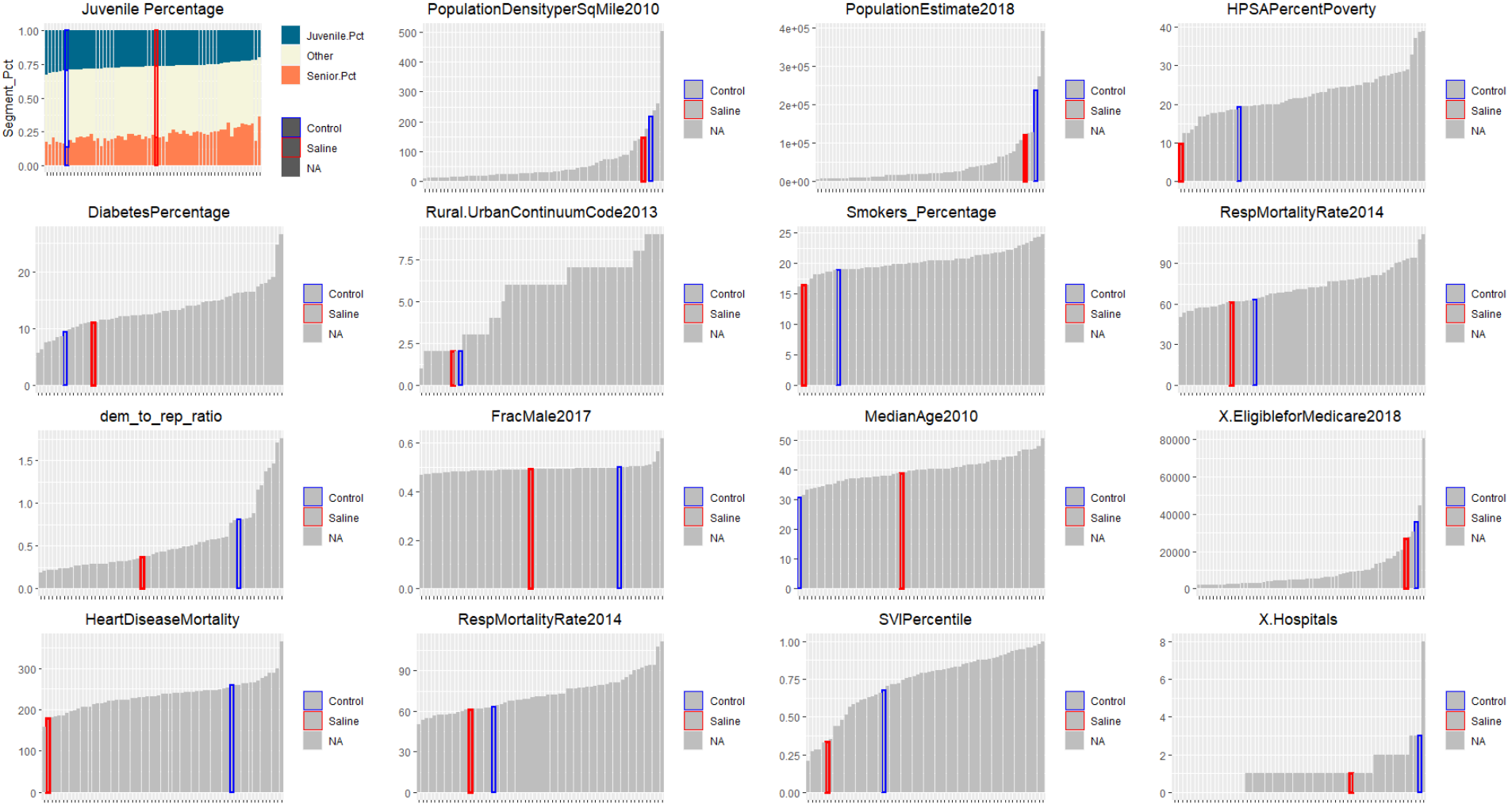}
	\caption{Covariate Values of Saline vs. Washington Counties}
	\label{covariates_saline_washington}
\end{figure}

\begin{figure}[H]
	\centering
	\includegraphics[width = 1\textwidth]{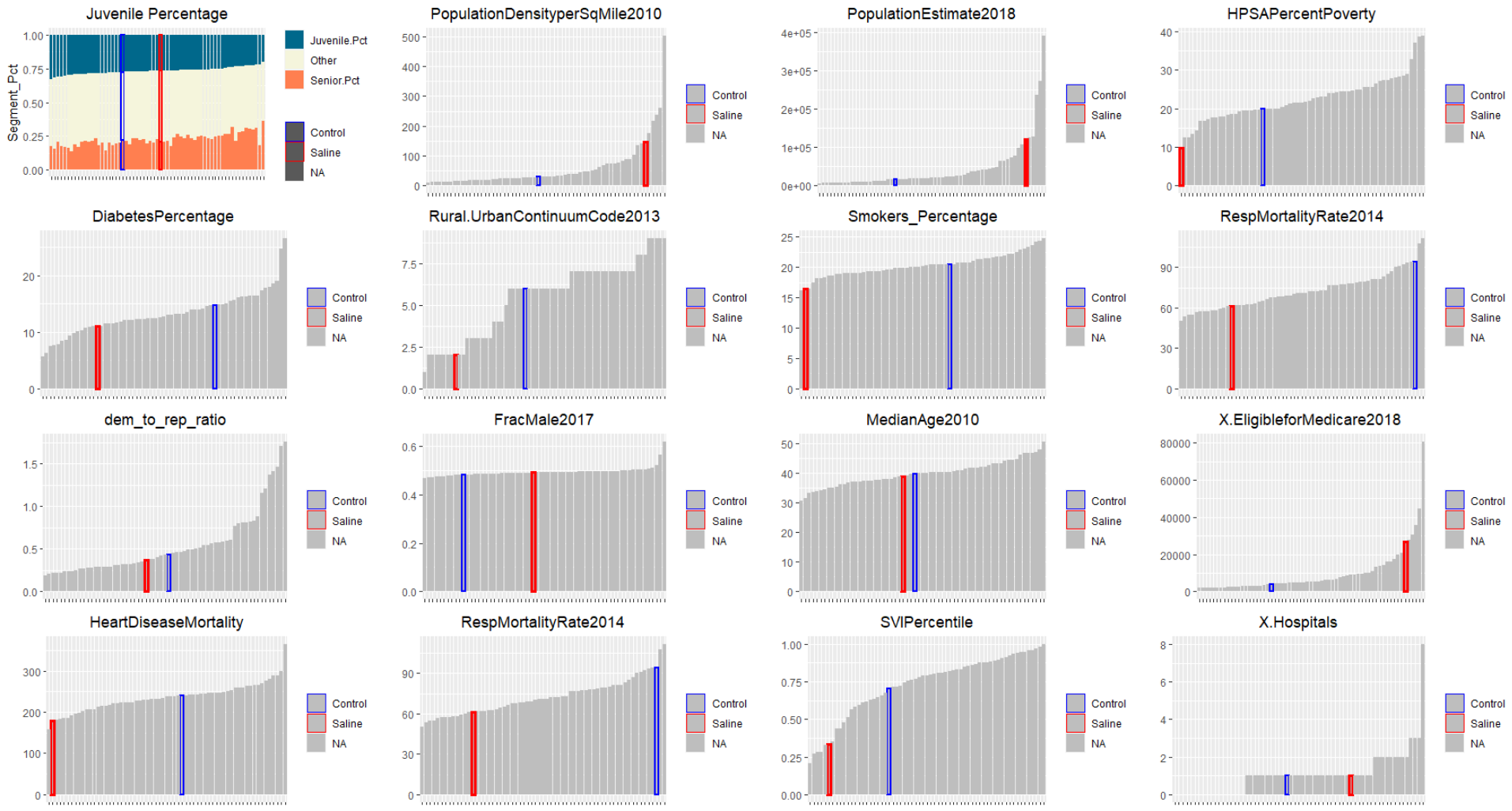}
	\caption{Covariate Values of Saline vs. Cross Counties}
	\label{covariates_saline_cross}
\end{figure}

\begin{figure}[H]
	\centering
	\includegraphics[width = 1\textwidth]{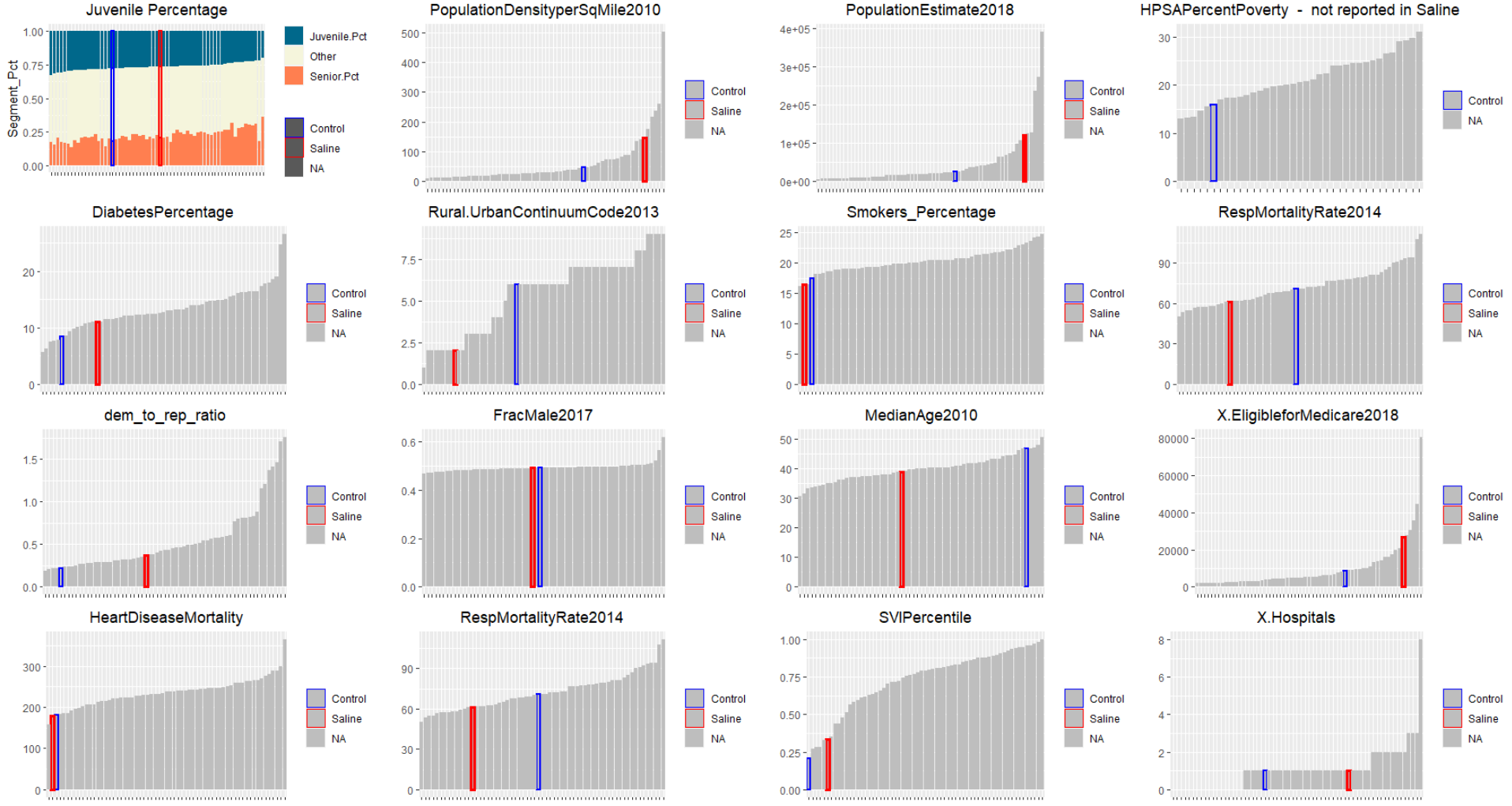}
	\caption{Covariate Values of Saline vs. Cleburne Counties}
	\label{covariates_saline_cleburne}
\end{figure}

\begin{figure}[H]
	\centering
	\includegraphics[width = 1\textwidth]{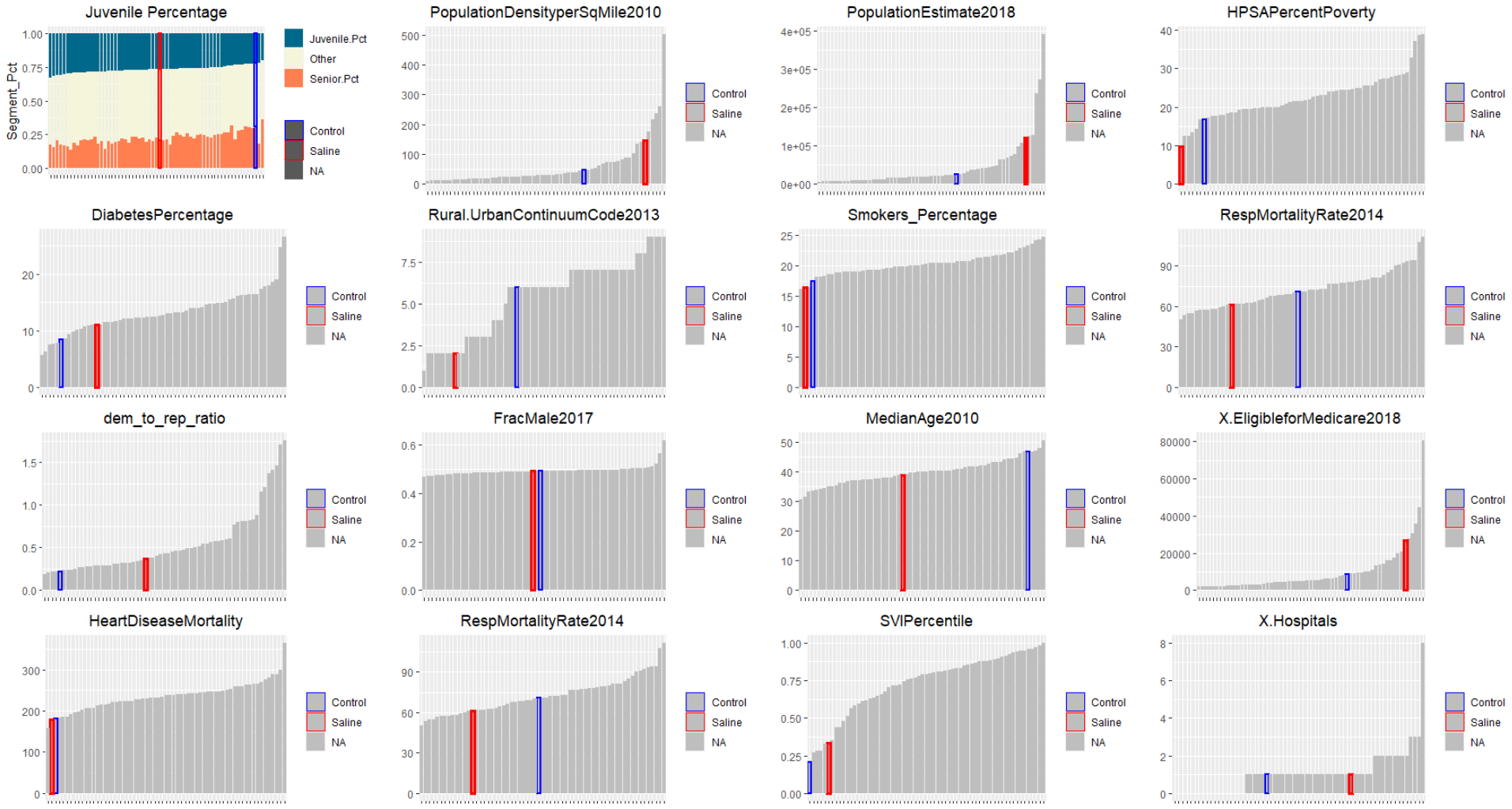}
	\caption{Covariate Values of Saline vs. Jefferson Counties}
	\label{covariates_saline_jefferson}
\end{figure}

\begin{figure}[H]
	\centering
	\includegraphics[width = 1\textwidth]{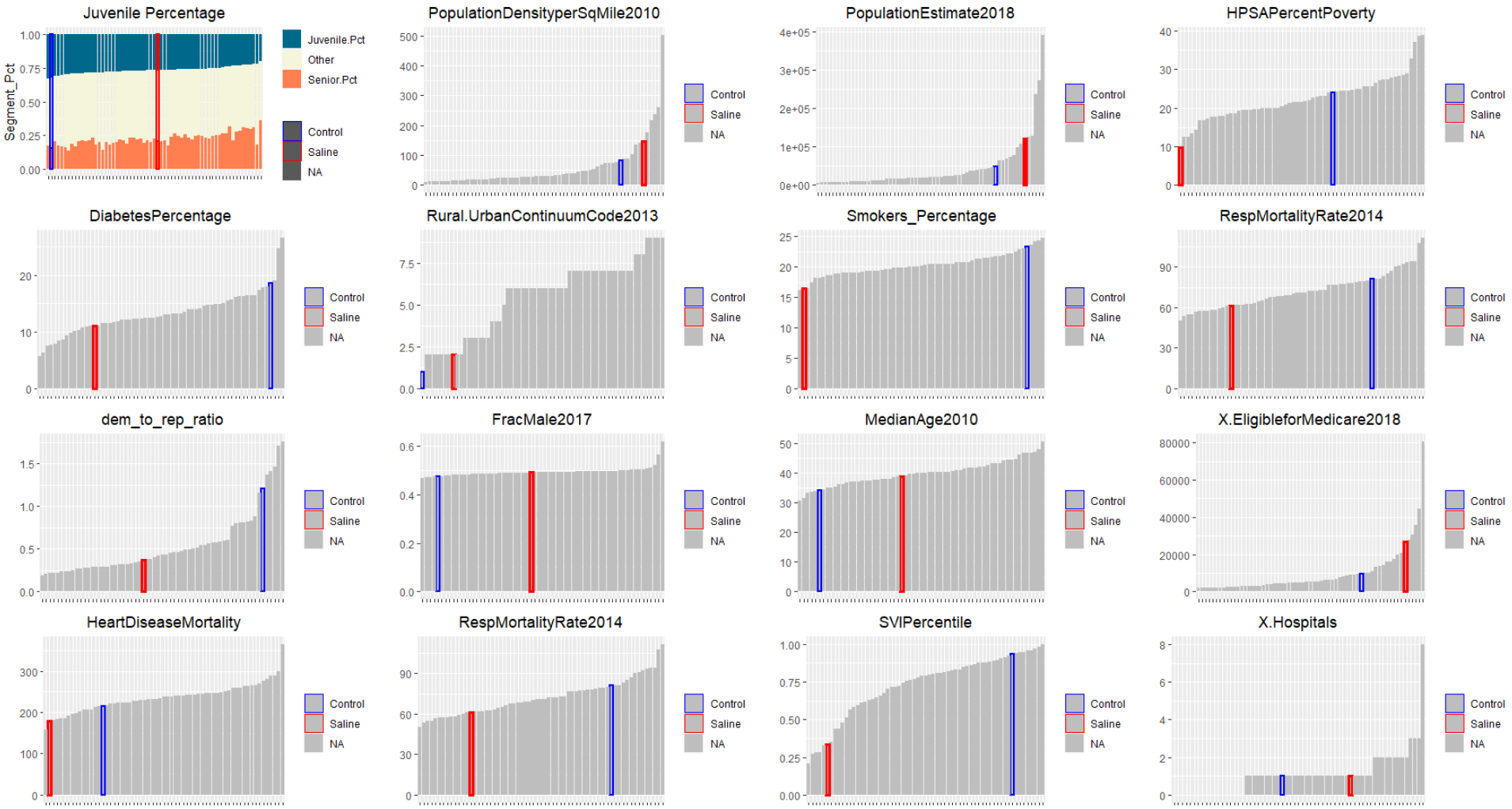}
	\caption{Covariate Values of Saline vs. Crittenden Counties}
	\label{covariates_saline_crittenden}
\end{figure}

\begin{figure}[H]
	\centering
	\includegraphics[width = 1.1\textwidth]{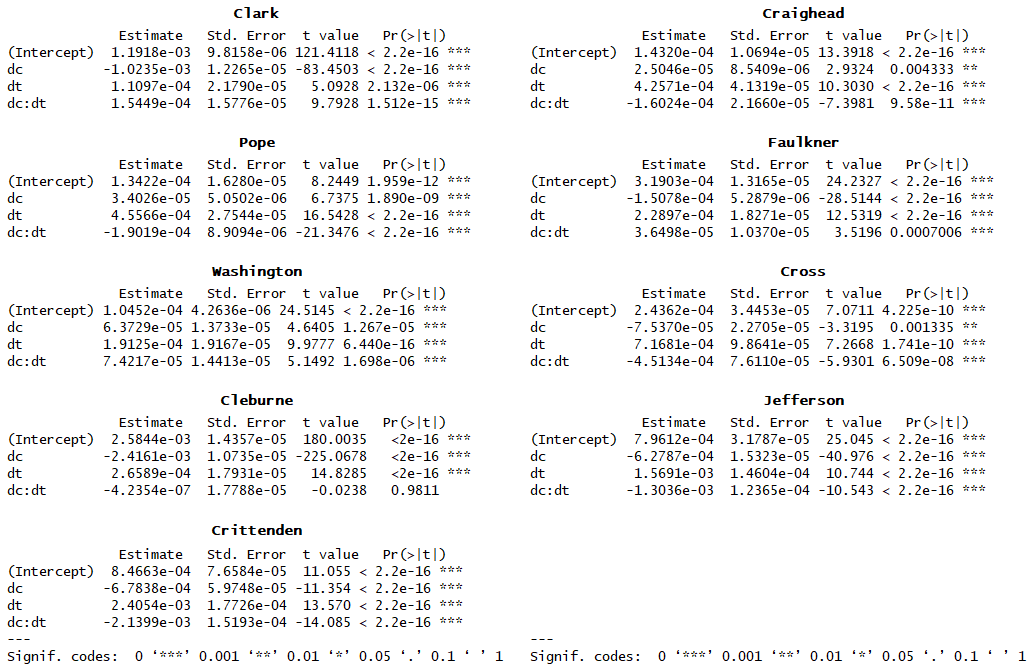}
	\caption{OLS Estimates of Treatment Effects with Clustered SEs}
	\label{lm_model_allcontrol}
\end{figure}

In Appendix, OLS output with clustered standard errors for all 11 control units in comparison to treatment unit (Saline county) is given in Figure \ref{lm_model_allcontrol}. Figure \ref{covariates_saline_clark}-\ref{covariates_saline_crittenden} shows how all the counties compare to the treatment unit (Saline county) along the covariates identified in Section \ref{SC}. Saline is more dense with lower mortality rates and residents in poverty, while Garland has a lower median age and more hospitals. 

\vspace{1cm}
\section{Synthetic Controls (SC)}\label{SC}
SC was proposed by Abadie et al. in a series of seminal papers [\citet{abadie2003economic, abadie2010synthetic, abadie2015comparative}] to estimate the counterfactual outcome of the treatment unit in absence of the treatment by using a weighted average of control units. Here, the weights of the control units are considered nuisance parameters that are estimated to arrive at the SC estimator. A useful quality that makes SC particularly apt for our current problem setup in Arkansas is that SC considers situations with one treatment and multiple control units. 

\subsection{Methodology}
Recall that $\textbf{X}_1$ represents the coefficients for the treatment unit for the covariates that are significant in explaining its \textit{pre-treatment }infection rates, and $\textbf{X}_0$ is a $K\times N$ matrix representing the values of the covariates for each of $N$ control units. The main task in the standard SC is to estimate the relative weights for the control units, called the \textit{SC weights}, by solving the following optimization problem [\cite{abadie2003economic}]:
\begin{equation}\label{SCoptimization}
	\textbf{W}^* = \min_{w\in \textbf{W}} (\textbf{X}_1 - \textbf{X}_0\textbf{W})'\textbf{V}(\textbf{X}_1-\textbf{X}_0\textbf{W})
\end{equation}
where $\textbf{W}=\{(w_1,...,w_N)\}$ are the weights of the $N$ control units, subject to $\sum_{i=1}^N w_i = 1$ and $w_i\ge 0$ for all $i\in [N]$. $\textbf{V} = Diag(v_1,...,v_K)$ where $v_i$ is the weight of the $i$-th covariate. Abadie et al. selected \textbf{V} such that $Y^1_{10}$ is best reproduced by SC $\textbf{W}^*(\textbf{V})$, where $\textbf{W}^*$ is the solution to \ref{SCoptimization}. 

Let $\textbf{Y}^1$ be a $T_0 \times 1$ vector representing the infection rates for Saline County where $T_0$ is the length of the pre-treatment period. Let $\textbf{Y}^0$ be a $T_0 \times N$ matrix containing the infection rates for $N$ potential control units. Then, 
\begin{equation}\label{SCoptimization2}
	\textbf{V}^* = \argmin_{v\in \mathbb{V}} (\textbf{Y}^1 -\textbf{ Y}_0\textbf{W}^*(\textbf{V}))'
	(\textbf{Y}^1 - \textbf{Y}_0\textbf{W}^*(\textbf{V}))
\end{equation}
where $\mathbb{V} := \{Diag(v_1,...,v_K) | v_i\ge 0\}$ and subject to the condition $\left\| \mathbb{V}\right\| = 1$ to ensure identifiability of the solution. Then, the SC weights are given by $\textbf{W}^*(\textbf{V}^*)$. 

Once the SC weights are computed, time series of the weighted average of the control group's infection rates and Saline County's infection rates are used to estimate the treatment effect, denoted by $\tau$ below:
\begin{equation}\label{SCregression}
	Y_{g_it}= \alpha + \beta'\cdot t + \gamma\cdot g_i + \tau \cdot g_i t +\varepsilon_{g_i}
\end{equation}
where $g_i\in \{1($Saline$), 0($Synthetic Control$)\}$.

\subsection{Covariate Selection}
The identification of meaningful covariates to explain variations in cases and deaths has been active research area since the inception of the pandemic. Wright et al. showed that low-income counties comply less with SIPO [\citet{wright2020poverty}]. Griffith et al. noted that men are more likely to become infected and have higher mortality due to biological, psychological, and behavioral factors [\citet{griffith2020men}]. Goldstein et al. found that the prevalence of the disease among 15-34-year-olds increased significantly faster than 34-49- amd 10-14-year-olds, suggesting the behavioral role in spreading the disease among the former [\citet{goldstein2020temporal}]. In addition, we consider other covariates in identifying \textbf{$X_1$}. Given reports in the literature about the incubation period of the virus [\citet{guan2020clinical, lauer2020incubation}] and the delay in policy effectiveness [\citet{abouk2020immediate}], a linear regression would not be a preferred tool of choice for assessing covariates. In addition, given the wide range of infection rates across the counties in Arkansas, we consider a more general negative binomial GLM that allows variance of the response to vary. 

\subsubsection{Negative Binomial GLM}\label{poissonglm}	
For $K_0$ potential covariates, let $\textbf{X}$ be a $K_0\times N$ matrix that contains the normalized values of those covariates for $N$ control units. Then, we run the following event study Negative Binomial GLM to select $K$ signficant covariates to base $X_1$ and $X_0$ in equation \ref{SCoptimization}. 
\begin{equation}\label{NBmodel}
	\log(\mathbb{E}[Y]) = \alpha_{0} + \beta' \textbf{X}
\end{equation}
with the variance of $Y$ given by $var(Y)=\mu + \mu^2/k$ where $\mu=\mathbb{E}[Y]$ and $k$ is the model dispersion parameter. The potential covariates considered included normalized values of the following in each county: males; those living below the poverty income threshold; juveniles; seniors (age 65 and over); population density per square mile; those with diabetics; county's CDC Social Vulnerability Index; number of full-time equivalent practitioners needed; those eligible for Medicare; ratio of voters who voted democratic versus republican; number of hospitals; the respiratory morality rate; and the heart disease mortality rate. Percentages of residents living below poverty threshold were not available for some of the counties on the CovidSeverity.com website. Hence, for poverty rates for all counties, we used the 2015 Arkansas Department of Health Report [\citet{arkansas_health}].

\subsubsection{Covariate Selection for SC}
Figure \ref{glmnb} shows the regression results of the model (\ref{NBmodel}):

\begin{figure}[H]
	\centering
	\includegraphics[width = .8\textwidth]{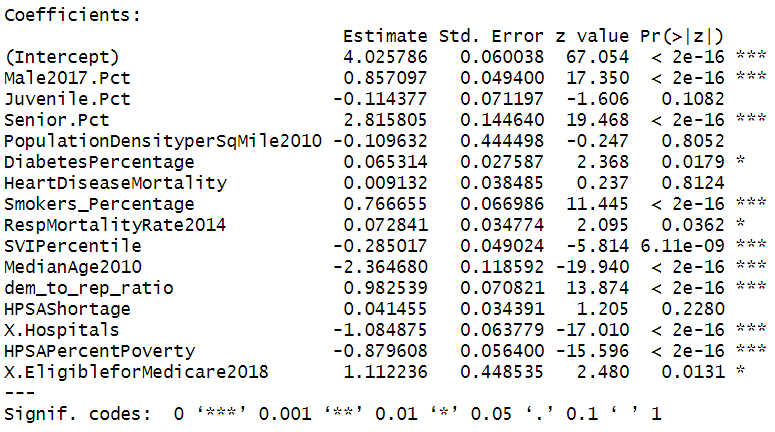}
	\caption{Negative Binomial GLM Output}
	\label{glmnb}
\end{figure}

\vspace{0.5cm}
As noted in the literature , men as percentage of the residents is significant [\citet{griffith2020men}], as is percentage of juveniles [\citet{goldstein2020temporal}]. Intuitively, seniors, population density, median age, and the number of hospitals in a county are significant. As noted in [\citet{wright2020poverty}], poverty rate is shown to be significant. We estimated $\textbf{V}$ in (\ref{SCoptimization}) using the magnitudes of the coefficient estimates in Figure \ref{glmnb} for use in (\ref{SCoptimization}). Then, we estimated the SC weights $\textbf{W}^*(\textbf{V}^*)$ using the iterative process involving  (\ref{SCoptimization}) and (\ref{SCoptimization2}). 

\subsubsection{Empirical Results}
\begin{figure}[H]
	\centering
	\includegraphics[width = .8\textwidth]{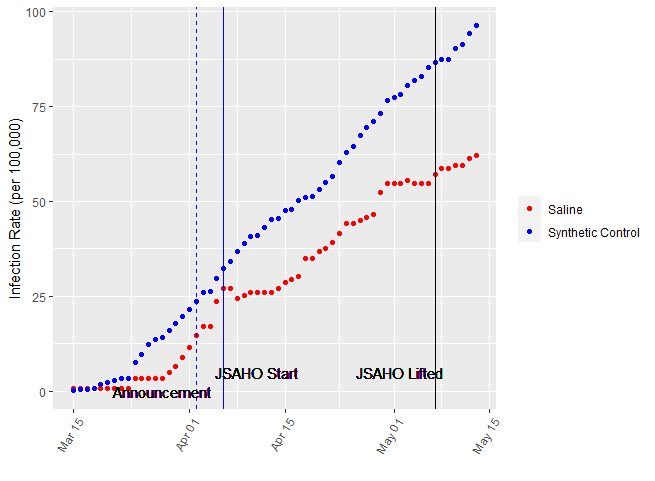}
	\caption{Infection Rates of Treatment and SC}
	\label{sc}
\end{figure}

Figure \ref{sc} shows the infection rates of Saline versus the Synthetic Control group weighted by $\textbf{W}^*(\textbf{V}^*)$. 

Below is the output from the regression output for (\ref{SCregression}):
\begin{figure}[H]
	\centering
	\includegraphics[width = .7\textwidth]{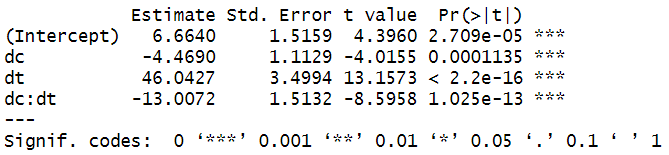}
	\caption{Estimate of Treatment Effect using SC}
	\label{sc_regression}
\end{figure}

\vspace{0.5cm}
As with the DID methodology, SC method yields the same conclusion.

\section{CONCLUSION}
There has been active multi-disciplinary research on Covid-19. However, to date little has been said about the causal impact of SAHO with limited scopes, such as JSAHO in Saline County, Arkansas. Using difference-in-differences and the synthetic controls design approaches, this paper presents evidence of a causal effect of county-level JSAHO implemented in a state that had not adopted a SAHO on reducing the growth rate of infection rates. While we studied the case in Saline County, the methods used here can be applied to assess the situations in other counties or local jurisdictions, and in the process strengthen the external validity of the findings by addressing the issues of limited duration and geographic specificity of the present study. There are other states that had not adopted statewide SIPO or SAHO when some of their local governments went ahead with their own orders at some point in the past, such as Utah. In addition, the analyses conducted in this paper can be applied to study the causal impact of other policy treatments. 

\section{DATA SOURCES}\label{data_sources}
Daily case counts for the counties in Arkansas were accessed on September 3, 2020, at \textit{The New York Times} Covid-19 site available at \url{https://github.com/nytimes/covid-19-data}. The data on county-level poverty rates were obtained from the Arkansas Department of Health report [\citet{arkansas_health}].  With the exception of the poverty rates data, data points for all other covariates were accessed on September 3, 2020, at the Covid-19 Severity Prediction project repository available at \url{http://covidseverity.com/}.

\subsubsection*{Acknowledgements}
We would like to thank Peter Bickel and Lihua Lei for helpful suggestions.

\newpage

\section{Appendix}
\subsection{Covariate Comparison Charts}

\vspace{1cm}

\vspace{1cm}

\vspace{1cm}

\vspace{1cm}

\vspace{1cm}

\vspace{1cm}

\vspace{1cm}

\vspace{1cm}

\vspace{1cm}

\vspace{1cm}

\newpage
\subsubsection{OLS Output for the Rest of the Candidate Control Units}

\newpage

\bibliography{xxx_journal}

\begin{thebibliography}{27}
\providecommand{\natexlab}[1]{#1}
\providecommand{\url}[1]{\texttt{#1}}
\expandafter\ifx\csname urlstyle\endcsname\relax
  \providecommand{\doi}[1]{doi: #1}\else
  \providecommand{\doi}{doi: \begingroup \urlstyle{rm}\Url}\fi

\bibitem[Abadie(2005)]{abadie2005semiparametric}
Alberto Abadie.
\newblock Semiparametric difference-in-differences estimators.
\newblock \emph{The Review of Economic Studies}, 72\penalty0 (1):\penalty0
  1--19, 2005.

\bibitem[Abadie and Gardeazabal(2003)]{abadie2003economic}
Alberto Abadie and Javier Gardeazabal.
\newblock The economic costs of conflict: A case study of the basque country.
\newblock \emph{American economic review}, 93\penalty0 (1):\penalty0 113--132,
  2003.

\bibitem[Abadie et~al.(2010)Abadie, Diamond, and
  Hainmueller]{abadie2010synthetic}
Alberto Abadie, Alexis Diamond, and Jens Hainmueller.
\newblock Synthetic control methods for comparative case studies: Estimating
  the effect of california's tobacco control program.
\newblock \emph{Journal of the American statistical Association}, 105\penalty0
  (490):\penalty0 493--505, 2010.

\bibitem[Abadie et~al.(2015)Abadie, Diamond, and
  Hainmueller]{abadie2015comparative}
Alberto Abadie, Alexis Diamond, and Jens Hainmueller.
\newblock Comparative politics and the synthetic control method.
\newblock \emph{American Journal of Political Science}, 59\penalty0
  (2):\penalty0 495--510, 2015.

\bibitem[Abouk and Heydari(2020)]{abouk2020immediate}
Rahi Abouk and Babak Heydari.
\newblock The immediate effect of covid-19 policies on social distancing
  behavior in the united states.
\newblock \emph{Available at SSRN}, 2020.

\bibitem[Ashenfelter and Card(1984)]{ashenfelter1984using}
Orley Ashenfelter and David Card.
\newblock Using the longitudinal structure of earnings to estimate the effect
  of training programs.
\newblock Technical report, National Bureau of Economic Research, 1984.

\bibitem[Bertrand et~al.(2004)Bertrand, Duflo, and
  Mullainathan]{bertrand2004much}
Marianne Bertrand, Esther Duflo, and Sendhil Mullainathan.
\newblock How much should we trust differences-in-differences estimates?
\newblock \emph{The Quarterly journal of economics}, 119\penalty0 (1):\penalty0
  249--275, 2004.

\bibitem[Caswell(2020{\natexlab{a}})]{caswell}
Bryn Caswell.
\newblock Stay-at-home order now in effect: What you need to know.
\newblock \emph{Dayton Now}, 2020{\natexlab{a}}.
\newblock URL
  \url{https://dayton247now.com/news/local/stay-at-home-order-goesinto-effect-at-midnight-what-you-need-to-know}.

\bibitem[Caswell(2020{\natexlab{b}})]{salinecountyorder}
Bryn Caswell.
\newblock Saline county judge issues juvenile stay at home' executive order.
\newblock \emph{Saline County}, 2020{\natexlab{b}}.
\newblock URL
  \url{https://www.salinecounty.org/plugins/show_image.php?id=2121}.

\bibitem[Chen et~al.(2020)Chen, Zhuo, de~la Fuente, Rohla, and
  Long]{chen2020causal}
M~Keith Chen, Yilin Zhuo, Malena de~la Fuente, Ryne Rohla, and Elisa~F Long.
\newblock Causal estimation of stay-at-home orders on sars-cov-2 transmission.
\newblock \emph{arXiv preprint arXiv:2005.05469}, 2020.

\bibitem[Chernozhukov et~al.(2020)Chernozhukov, Kasaha, and
  Schrimpf]{chernozhukov2020causal}
Victor Chernozhukov, Hiroyuki Kasaha, and Paul Schrimpf.
\newblock Causal impact of masks, policies, behavior on early covid-19 pandemic
  in the us.
\newblock \emph{arXiv preprint arXiv:2005.14168}, 2020.

\bibitem[Courtemanche et~al.(2020)Courtemanche, Garuccio, Le, Pinkston, and
  Yelowitz]{courtemanche2020strong}
Charles Courtemanche, Joseph Garuccio, Anh Le, Joshua Pinkston, and Aaron
  Yelowitz.
\newblock Strong social distancing measures in the united states reduced the
  covid-19 growth rate: Study evaluates the impact of social distancing
  measures on the growth rate of confirmed covid-19 cases across the united
  states.
\newblock \emph{Health Affairs}, pages 10--1377, 2020.

\bibitem[Dave et~al.(2020)Dave, Friedson, Matsuzawa, and
  Sabia]{dave2020shelter}
Dhaval Dave, Andrew~I Friedson, Kyutaro Matsuzawa, and Joseph~J Sabia.
\newblock When do shelter-in-place orders fight covid-19 best? policy
  heterogeneity across states and adoption time.
\newblock \emph{Economic Inquiry}, 2020.

\bibitem[DoH(2015)]{arkansas_health}
Arkansas DoH.
\newblock Arkansas department of health office of rural health and primary care
  primary care needs assessment.
\newblock \emph{Arkansas Department of Health}, 2015.
\newblock URL
  \url{https://www.healthy.arkansas.gov/images/uploads/pdf/PCO_Needs_Assessment_September_Revision_CoverPageEdit.pdf}.

\bibitem[Friedson et~al.(2020)Friedson, McNichols, Sabia, and
  Dave]{friedson2020did}
Andrew~I Friedson, Drew McNichols, Joseph~J Sabia, and Dhaval Dave.
\newblock Did california's shelter-in-place order work? early
  coronavirus-related public health effects.
\newblock Technical report, National Bureau of Economic Research, 2020.

\bibitem[Gao et~al.(2020)Gao, Rao, Kang, Liang, Kruse, Dopfer, Sethi, Reyes,
  Yandell, and Patz]{gao2020association}
Song Gao, Jinmeng Rao, Yuhao Kang, Yunlei Liang, Jake Kruse, Dorte Dopfer,
  Ajay~K Sethi, Juan Francisco~Mandujano Reyes, Brian~S Yandell, and Jonathan~A
  Patz.
\newblock Association of mobile phone location data indications of travel and
  stay-at-home mandates with covid-19 infection rates in the us.
\newblock \emph{JAMA Network Open}, 3\penalty0 (9):\penalty0
  e2020485--e2020485, 2020.

\bibitem[Goepp et~al.(2018)Goepp, Bouaziz, and Nuel]{goepp2018spline}
Vivien Goepp, Olivier Bouaziz, and Gr{\'e}gory Nuel.
\newblock Spline regression with automatic knot selection.
\newblock \emph{arXiv preprint arXiv:1808.01770}, 2018.

\bibitem[Goldstein and Lipsitch(2020)]{goldstein2020temporal}
Edward Goldstein and Marc Lipsitch.
\newblock Temporal rise in the proportion of younger adults and older
  adolescents among coronavirus disease (covid-19) cases following the
  introduction of physical distancing measures, germany, march to april 2020.
\newblock \emph{Eurosurveillance}, 25\penalty0 (17):\penalty0 2000596, 2020.

\bibitem[Griffith et~al.(2020)Griffith, Sharma, Holliday, Enyia, Valliere,
  Semlow, Stewart, and Blumenthal]{griffith2020men}
Derek~M Griffith, Garima Sharma, Christopher~S Holliday, Okechuku~K Enyia,
  Matthew Valliere, Andrea~R Semlow, Elizabeth~C Stewart, and Roger~Scott
  Blumenthal.
\newblock Men and covid-19: A biopsychosocial approach to understanding sex
  differences in mortality and recommendations for practice and policy
  interventions.
\newblock \emph{Preventing chronic disease}, 17:\penalty0 E63, 2020.

\bibitem[Guan et~al.(2020)Guan, Ni, Hu, Liang, Ou, He, Liu, Shan, Lei, Hui,
  et~al.]{guan2020clinical}
Wei-jie Guan, Zheng-yi Ni, Yu~Hu, Wen-hua Liang, Chun-quan Ou, Jian-xing He,
  Lei Liu, Hong Shan, Chun-liang Lei, David~SC Hui, et~al.
\newblock Clinical characteristics of coronavirus disease 2019 in china.
\newblock \emph{New England journal of medicine}, 382\penalty0 (18):\penalty0
  1708--1720, 2020.

\bibitem[Hsiang et~al.(2020)Hsiang, Allen, Annan-Phan, Bell, Bolliger, Chong,
  Druckenmiller, Huang, Hultgren, Krasovich, et~al.]{hsiang2020effect}
Solomon Hsiang, Daniel Allen, S{\'e}bastien Annan-Phan, Kendon Bell, Ian
  Bolliger, Trinetta Chong, Hannah Druckenmiller, Luna~Yue Huang, Andrew
  Hultgren, Emma Krasovich, et~al.
\newblock The effect of large-scale anti-contagion policies on the covid-19
  pandemic.
\newblock \emph{Nature}, 584\penalty0 (7820):\penalty0 262--267, 2020.

\bibitem[Lauer et~al.(2020)Lauer, Grantz, Bi, Jones, Zheng, Meredith, Azman,
  Reich, and Lessler]{lauer2020incubation}
Stephen~A Lauer, Kyra~H Grantz, Qifang Bi, Forrest~K Jones, Qulu Zheng,
  Hannah~R Meredith, Andrew~S Azman, Nicholas~G Reich, and Justin Lessler.
\newblock The incubation period of coronavirus disease 2019 (covid-19) from
  publicly reported confirmed cases: estimation and application.
\newblock \emph{Annals of internal medicine}, 172\penalty0 (9):\penalty0
  577--582, 2020.

\bibitem[Le et~al.(2020)Le, Le, Brooks, Khetpal, Liauw, Izurieta, and
  Reina~Ortiz]{le2020impact}
NK~Le, AV~Le, JP~Brooks, S~Khetpal, D~Liauw, R~Izurieta, and M~Reina~Ortiz.
\newblock Impact of government-imposed social distancing measures on covid-19
  morbidity and mortality around the world.
\newblock \emph{Bull World Health Organ}, 10, 2020.

\bibitem[Lurie et~al.(2020)Lurie, Silva, Yorlets, Tao, and
  Chan]{lurie2020covid}
Mark~N Lurie, Joe Silva, Rachel~R Yorlets, Jun Tao, and Philip~A Chan.
\newblock Covid-19 epidemic doubling time in the united states before and
  during stay-at-home restrictions.
\newblock \emph{The Journal of Infectious Diseases}, 2020.

\bibitem[Napoleon(2020)]{napoleon_2020}
Carrie Napoleon.
\newblock "police hoping public adheres to holcomb's shelter-in-place order,
  but will be vigilant about enforcement".
\newblock \emph{chicago tribune}, Mar 2020.
\newblock URL
  \url{https://www.chicagotribune.com/suburbs/post-tribune/ct-ptb-police-holcomb-enforcement-st-0326-20200331-
  cdng7tjsb5gm5asmd3jh5pttjy-story.html}.

\bibitem[Ronayne and Thompson(2020)]{ronayne}
Kathleen Ronayne and Don Thompson.
\newblock California governor issues statewide stay-at home order.
\newblock \emph{AP News}, 2020.
\newblock URL \url{https://apnews.com/9ca4a191790dd6f80bd5acec569ec423}.

\bibitem[Wright et~al.(2020)Wright, Sonin, Driscoll, and
  Wilson]{wright2020poverty}
Austin~L Wright, Konstantin Sonin, Jesse Driscoll, and Jarnickae Wilson.
\newblock Poverty and economic dislocation reduce compliance with covid-19
  shelter-in-place protocols.
\newblock \emph{University of Chicago, Becker Friedman Institute for Economics
  Working Paper}, 2020.

\end{thebibliography}
\bibliographystyle{plainnat}

\end{document}